\documentclass[%
 reprint,
nofootinbib,
 amsmath,amssymb,
 aps,
floatfix,
]{revtex4-2}

\usepackage{graphicx}
\usepackage{dcolumn}
\usepackage{bm}
\usepackage{siunitx}

\newcommand{\FNAL}{\affiliation{Fermi National Accelerator Laboratory, Batavia, IL 60510, USA}}
\newcommand{\IIT}{\affiliation{Illinois Institute of Technology, Chicago, IL 60616, USA}}

\begin{document}

\preprint{APS/123-QED}

\title{Precision Beam Diagnostics at the NuMI Facility Using Muon Monitor Observations}

\author{Katsuya Yonehara}
 \email{yonehara@fnal.gov}\FNAL
\author{Sudeshna Ganguly}\FNAL
 \author{Don Athula Wickremasinghe}\FNAL
\author{Pavel Snopok}\IIT
 \author{Yiding Yu}\IIT

\date{\today}

\begin{abstract}
The Neutrinos at the Main Injector (NuMI) facility at Fermilab delivers an intense neutrino beam for multiple experiments by producing pions that decay into neutrinos, muons, and other particles. 
Magnetic horns---the primary pion focusing elements in the NuMI beamline---exhibit predominantly linear optics, enabling a predictable relationship between the proton beam and the resulting pion and muon phase spaces. 
This study has two primary objectives: first, to evaluate and confirm the linearity of the horn focusing mechanism using analytical models and numerical simulations; and second, to demonstrate that key beam parameters---such as proton beam intensity, beam position on target, and horn current---can be extracted from muon monitor observations within this linear optics framework.
Our results show that muon beam profiles can detect horn current variations with a $\pm 0.05\%$ precision. 
This approach provides a reliable cross-check of beam parameters, helping to reduce systematic uncertainties that are critical for future experiments such as the Deep Underground Neutrino Experiment (DUNE), which will rely on the neutrino beam produced by the Long Baseline Neutrino Facility (LBNF).

\end{abstract}

\maketitle


\section{Introduction}
Fermilab hosts the most powerful accelerator-produced neutrino beam facility in the world, enabling precision studies of neutrino properties. 
In the Neutrinos at the Main Injector (NuMI) facility~\cite{adamson2016}, an intense proton beam strikes a high-power target, generating positively and negatively charged pions through hadronic interactions. 
The primary pion decay processes for neutrino production are: 
\begin{align*}
    \pi^+ \rightarrow \nu_\mu + \mu^+, \\
    \pi^- \rightarrow \bar{\nu}_\mu + \mu^-.
\end{align*}
A magnetic horn plays a crucial role in the neutrino beamline by selecting pions of the appropriate charge and energy and focusing them toward a decay volume. 
The subsequent decays produce neutrinos with the desired flavor and chirality, as required by the weak interaction.
\footnote{Only left-handed neutrinos and right-handed antineutrinos participate in weak interactions. Right-handed neutrinos are not part of the Standard Model and are considered sterile.}
Horn magnets with specific conductor shapes and configurations have been developed for various neutrino experiments~\cite{Charitonidis, Helmut,  ICHIKAWA, KOPP_2007, Palmer, VandeerMeer}, based on the expectation that they would exhibit approximately linear optics.
Currently, the NuMI neutrino beam---covering neutrino energies from below $1$\,GeV to approximately $20$\,GeV---is delivered to the NOvA~\cite{NOvA:2019cyt,NOvA:2016kwd} and ICARUS~\cite{icarus,ICARUS:2022huv} detectors; it was also previously used by the MicroBooNE experiment~\cite{MicroBooNE:2021gfj,MicroBooNE:2016pwy}.
Given its crucial role in pion focusing, the horn must produce a stable and reliable magnetic field with minimal systematic uncertainty. 
The allowable beam-related systematic uncertainty for neutrino measurements at the NOvA detectors has been quantitatively estimated~\cite{Cremonesi2017}. 
The corresponding beam instrumentation used to monitor these parameters is listed in Table~\ref{tab:phys}.
\begin{table*}
\caption{\label{tab:phys}
Estimated physics tolerance for the beam parameters of the NOvA experiment and the corresponding beam instrumentation. 
Here, CT is a current transformer, BPM is a beam position monitor, and PM is a beam profile monitor. POT/spill stands for protons on target per single beam spill. 
Accuracy refers to the absolute calibration uncertainty; stability indicates the pulse-to-pulse variation of the measurement.}
\begin{ruledtabular}
\begin{tabular}{ lcclc } 
Beam parameter & Design value & Tolerance & Beam instrument & Accuracy \\ \hline 
Horn current & $200$\,kA & $\pm2$\,kA & Horn CT & $\pm0.1$\% \\ 
 Horiz. beam position on target  & $0$\,mm & $\pm1$\,mm & BPM & $\pm0.02$\,mm\\
 Vertical beam position on target & $0$\,mm & $\pm1$\,mm & BPM & $\pm0.02$\,mm\\
 RMS beam spot size & $1.3$\,mm & $\pm0.2$\,mm & Beam PM & $\pm0.1$\,mm\\
 Beam intensity & $50\times10^{12}$\,POT/spill & $1$\% & Beam CT & $\pm0.5$\% (stability $\pm0.1$\%)\\
\end{tabular}
\end{ruledtabular}
\end{table*}
The primary beam parameters are monitored using conventional detectors, with their measurement accuracies listed in Table~\ref{tab:phys}. 
However, fluctuations in the primary beam detector signals cannot be independently verified, limiting confidence in their accuracy. 
To provide an independent cross-check, the NuMI beamline employs three muon monitors located downstream of the decay pipe and hadron absorber (see Fig.~\ref{fig:schem}). 
Each monitor measures the muon beam profile resulting from pions focused by the horn magnet and subsequently decaying along the NuMI beamline. 
As illustrated in Fig.~\ref{fig:schem}, the total absorber material---including the hadron absorber and the intervening rock---varies in thickness in front of and between the monitors. 
As a result, each muon monitor samples a distinct portion of the muon energy spectra. 
Consequently, the resulting muon beam profiles reflect the horn's energy-dependent focusing behavior, commonly referred to as chromaticity. 

Here, we examine the horn focusing mechanism to understand how variations in horn current, beam intensity, and beam position at the target affect the observed muon beam profiles. 
An analytical model of a single horn is developed and extended to a semi-analytical model of the full multi-horn system, demonstrating that the NuMI horn system exhibits predominantly linear beam optics. 
These models are validated using particle tracking simulations to confirm their accuracy in describing the horn's focusing behavior.  
Furthermore, a machine learning model is employed to analyze variations in the muon beam profiles and extract key beam parameters with precision comparable to that of the primary beam instrumentation. 
As a result, the horn current is reconstructed with a precision of $\pm0.05$\%, the beam intensity with $\pm0.1$\%, and the proton beam position at the target with $\pm0.018$\,mm horizontally and $\pm0.013$\,mm vertically, based on muon monitor observations. 

The structure of the paper is as follows: Sec.~\ref{sec:numibeamcomponents} introduces the NuMI beamline;  Sec.~\ref{sec:hornfocusingmechanism} presents analytical and numerical investigations of the horn focusing mechanism;  
Sec.~\ref{sec:beamresponsemeasurement} describes the observed beam response; 
Sec.~\ref{sec:machinelearning} details the machine learning approach; 
and Sec.~\ref{sec:summary} concludes with a summary and future outlook. 

\section{NuMI beamline and Instrumentation}
\label{sec:numibeamcomponents}
\begin{figure*}
\includegraphics[width=\linewidth]{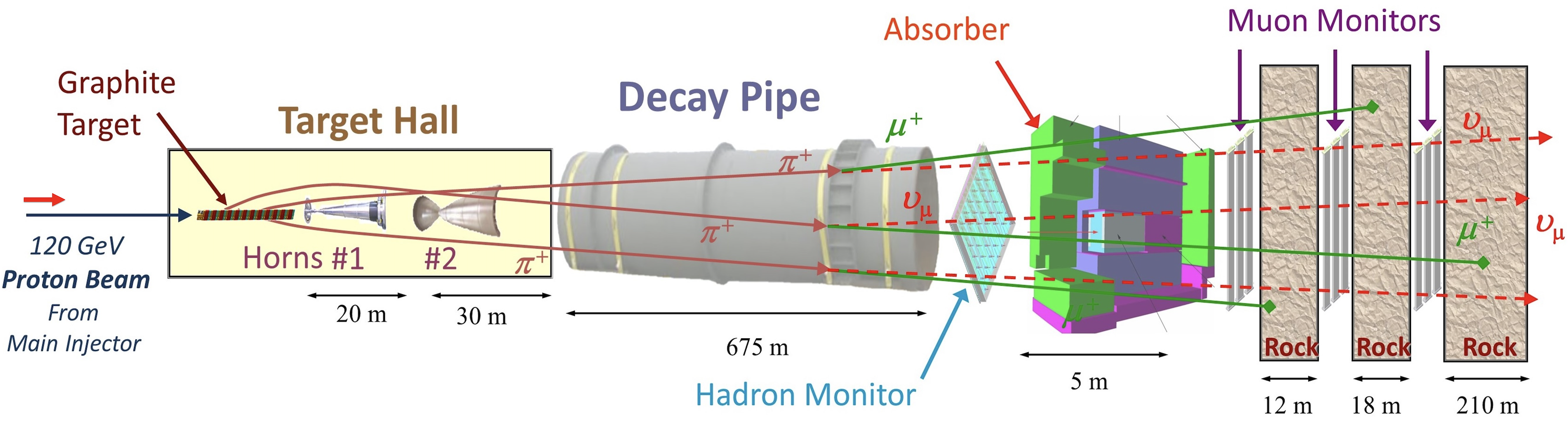}
\caption{\label{fig:schem}
The overall layout of the NuMI beamline.}
\end{figure*}

Operational since 2005, the NuMI facility has undergone several upgrades, most notably a 2017 enhancement under the Accelerator Improvement Plan (AIP)~\cite{Convery2017}, which allowed the Fermilab accelerator complex to achieve a beam power of 1 megawatt (MW). 
Figure~\ref{fig:schem} illustrates the current configuration of the NuMI target and horn system, together with the associated muon monitors. 
A 120-GeV/$c$ proton beam is delivered from the Main Injector to the NuMI target system every $1.067$ seconds. 
The beam spill length is \qty{9.6}{\us}. 
The achievable beam intensity is $5.6\times10^{13}$ protons on target (POT) per spill.
The structure of the proton beam in the Main Injector and the optics of the NuMI proton beam transport line are described in~\cite{adamson2016}.
Minor updates have been made to adjust the RMS beam spot size at the target to approximately 1.5\,mm for nominal operation in 2024~\cite{yonehara2024}. 
The following subsections describe key beam line components and instrumentation used to study muon monitor profiles. 
\subsection{Key NuMI Beamline Components}
\begin{figure*}
\includegraphics[width=\linewidth]{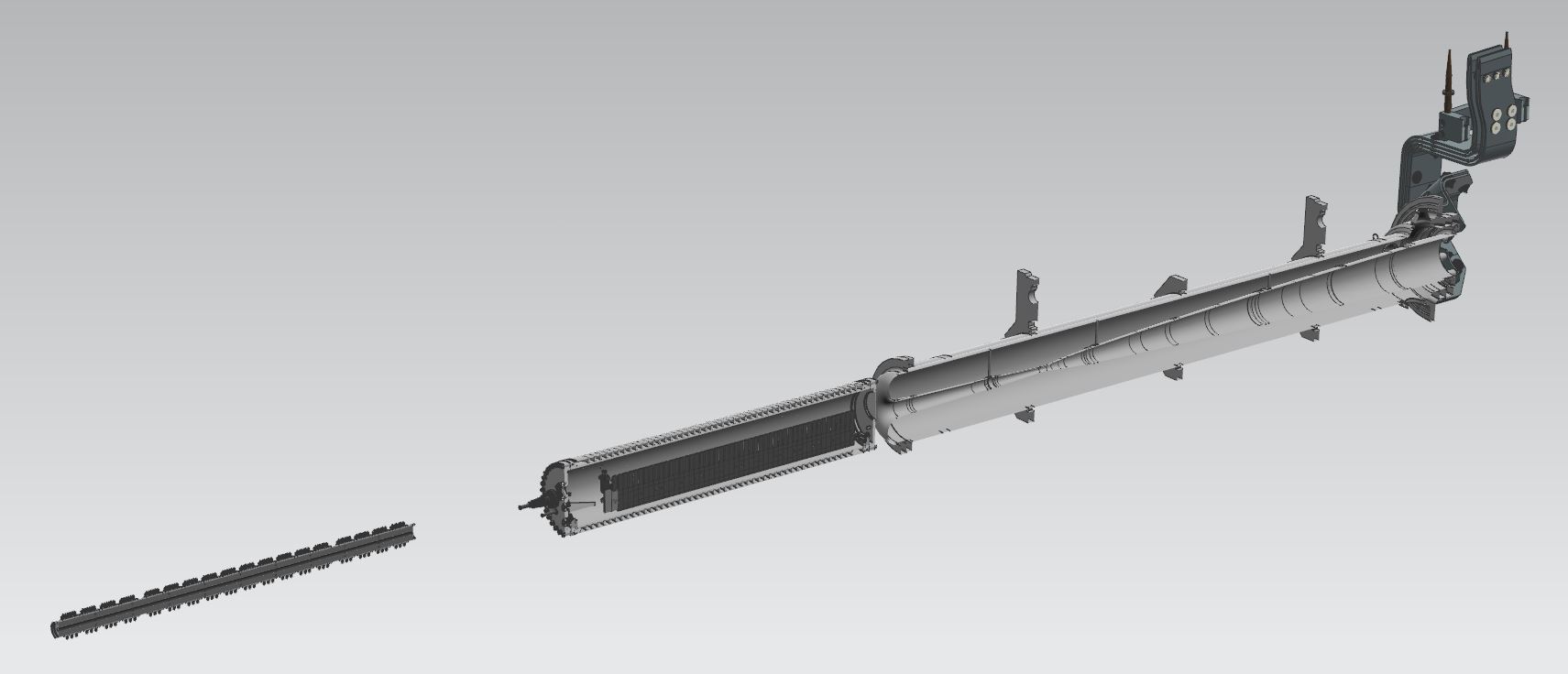}
\caption{\label{fig:layout1}
Cross-sectional view of the baffle, target canister, and Horn~1. 
The proton beam enters from the bottom left and exits toward the upper right.}
\end{figure*}

Figure~\ref{fig:layout1} shows the layout of the baffle, target canister, and Horn~1. 
The baffle is a 1.5-meter-long graphite cylinder with a 15-mm-diameter central aperture designed to intercept mis-steered proton beams. 
The target canister is a water-cooled aluminum cylinder filled with atmospheric pressure helium gas that houses the target core. 
Thin beryllium foils serve as the upstream and downstream beam windows, with the downstream window measuring $120$\,mm in diameter and $1.25$\,mm in thickness. 
Both the baffle and target canister are mounted on the target module. 
The distances between the baffle and the upstream end of the target core, and between the downstream end of the target core and Horn~1, are approximately 870\,mm and $200$\,mm, respectively. 

\begin{figure}[htb]
\includegraphics[width=8.5cm]{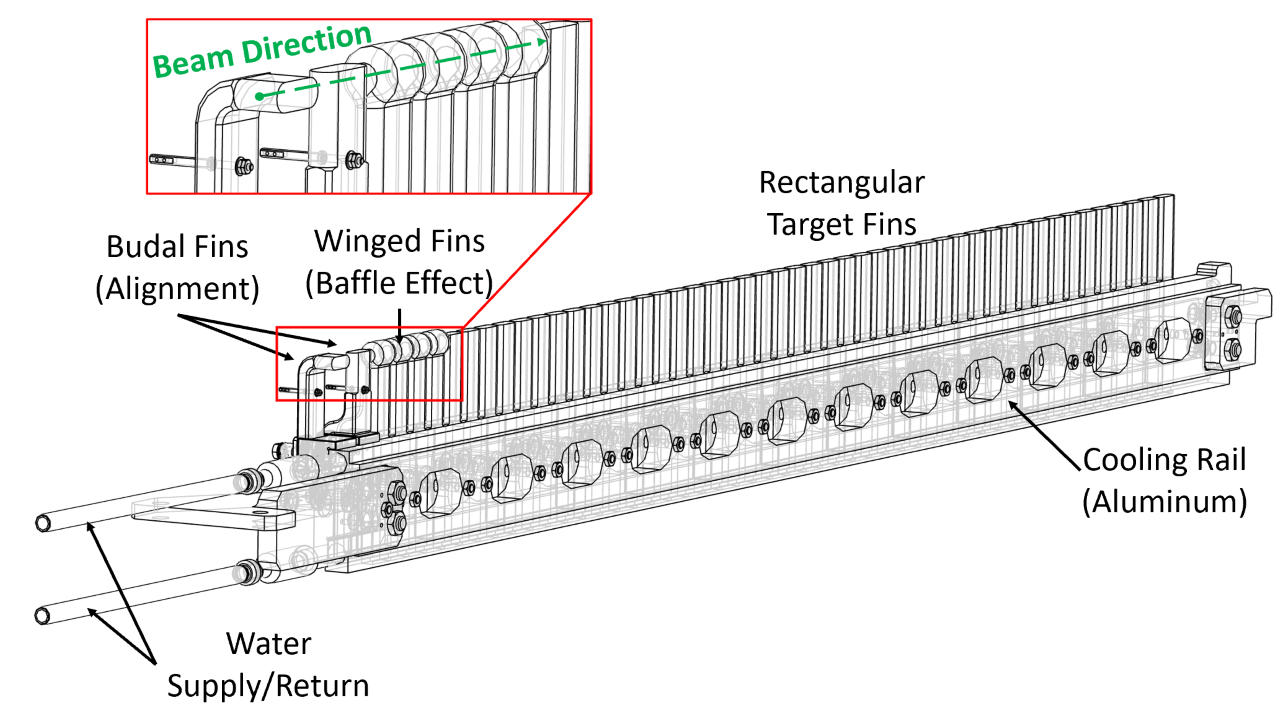}
\caption{\label{fig:targetfin}
Layout of the NuMI target core, including the Budal fins, winged fins, and regular target fins.}
\end{figure}

Figure~\ref{fig:targetfin} shows the target core made of a series of graphite fins (grade: POCO ZXF-5Q, density: $1.78$ g/cm$^3$). 
The nominal beam position at the target core is near the upper end of the fin. 
The first two fins, known as Budal fins, are used for coarse beam position measurements~\cite{adamson2016}. 
The next four fins, called winged fins, feature a thick graphite cylinder around the top of the fin. 
The diameter of the cylinder, $18$\,mm exceeds the baffle's central aperture, enabling the interception of mis-steered beams and mitigation of their thermal impact on downstream components through multiple scattering. 
The rest of the target core has 44 rounded rectangular target fins, each measuring $25$\,mm in length and $9$\,mm in width. 
The total length of the target core is $1.25$\,m, corresponding to approximately $2.5$\,nuclear interaction lengths for the proton beam. 

\begin{figure*}
\includegraphics[width=16.2cm]{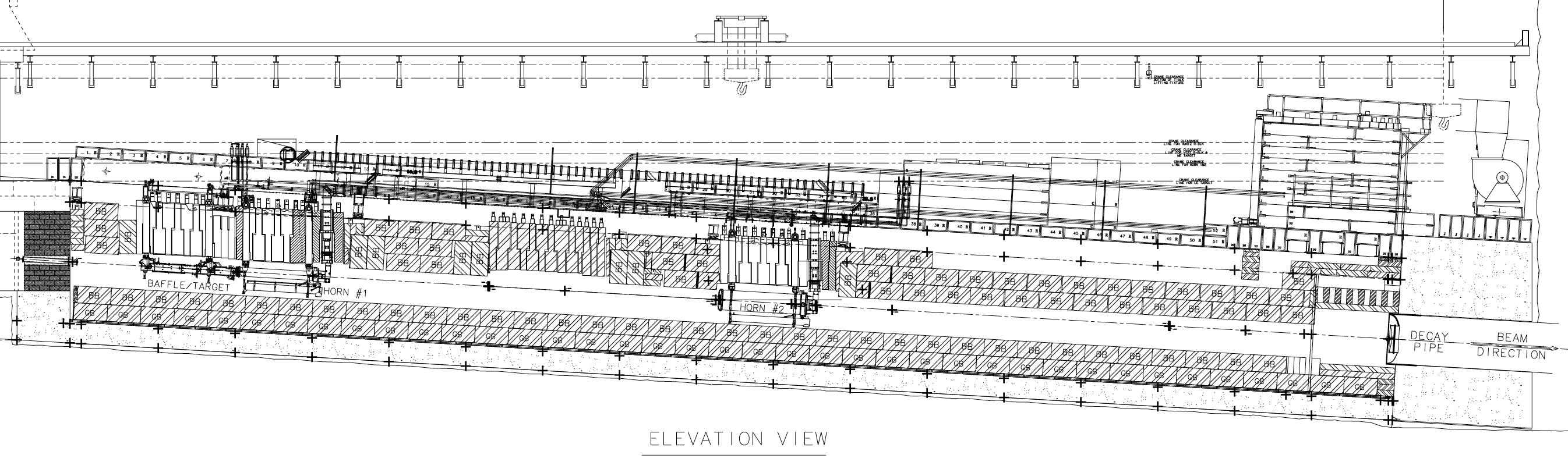}
\caption{\label{fig:layout2}
The elevation view of the target hall.
}
\end{figure*}

Figure~\ref{fig:layout2} shows an elevation view of the target hall, including the target module, Horn~1, Horn~2, and the upstream end of the decay pipe. 
The beam axis is tilted downward by $58$\,mrad to direct the beam toward the Soudan Mine in Minnesota, the site of the former MINOS far detector~\cite{Michael2008}. 
The distance between Horn~1 and Horn~2 is $20$\,m, and the distance from Horn~2 to the upstream end of the decay pipe is $25$\,m. 
Figure~\ref{fig:horncurrent} shows the cross-sectional view of Horn~1. 
The inner conductor has a parabolic shape at both ends, connected by a narrow cylindrical ``neck'' located at the parabolic vertices, 
while the outer conductor is a straight cylinder. 
A curved transition region at the upstream end connects the two conductors, and a stripline is attached at the downstream end for current injection and extraction. 
Horn~2 has a similar geometry, though with a larger radial size; detailed dimensions are provided in Sec.~\ref{sec:semi-analytical_model}.
Horns~1 and~2 are electrically connected in series. 



A Forward Horn Current (FHC) mode---commonly known as neutrino mode---occurs when the horn current flows in the inner conductor in the same direction as the proton beam through the inner conductor. 
This generates a clockwise toroidal magnetic field between the inner and outer conductors, which focuses positively charged pions, as illustrated in Fig.~\ref{fig:horncurrent}. 
Conversely, in the Reverse Horn Current (RHC) mode, or antineutrino mode, the current flows in the opposite direction, focusing negative charged pions.
\begin{figure*}
\includegraphics[width=\linewidth]{Figures/horn1currentflow.png}
\caption{\label{fig:horncurrent}
Cross-sectional view of Horn~1 showing the parabolic inner conductor enclosed by the outer conductor. 
A red arrow indicates the direction of the horn current in forward horn current (FHC) mode. 
Colored trajectories represent pions with different momenta: red ($5$\,GeV/$c$), green ($10$\,GeV/$c$), and blue ($20$\,GeV/$c$). 
A wrong-sign pion (black, $10$\,GeV/$c$) is defocused. 
}
\end{figure*}

To verify the agreement with the design specifications, the radial magnetic field strength at the horn neck was measured using a Hall probe before beam exposure. 
The measured radial distance between the outer conductor and the outer surface of the neck was within the design tolerance of $\pm0.005$ inches. 
Using these measured values as input, the calculated magnetic field strength agreed with analytical predictions and design specifications to within $\pm0.1\%$~\cite{Hylen2009}.


The inner conductor is cooled by water jets sprayed symmetrically from the outer conductor at the 12, 4, and 8 o'clock positions.  
In simulations, the total effective thickness of the water layer is assumed to be $1$\,mm. 
The volume between the inner and outer conductors is filled with atmospheric pressure argon gas to displace air, which would otherwise lead to the formation of corrosive nitric acid. 
Argon is chosen over helium because of its higher electrical breakdown voltage and lower cost.

The decay pipe is filled with atmospheric pressure helium gas, where most pions decay. 
The pipe has a length of $675$\,m and an inner diameter of $2$\,m. 
Aluminum windows with thicknesses of $1.6$ and $6.25$\,mm are installed at the upstream and downstream ends, respectively. 
The first beam monitor downstream of the decay pipe is the hadron monitor. 
It is a grid ionization chamber that is used to measure the spatial distribution of remnant protons and hadrons~\cite{Zwaska2006}. 
Due to prolonged exposure to high-intensity charged particle flux, the hadron monitor gradually experiences signal degradation from radiation damage. 
Therefore, it is used primarily for proton beam alignment~\cite{adamson2016,Zwaska:2006px} rather than for routine beam quality monitoring. 
The positions of the target and horns are determined through proton beam tomography, which utilizes multiple selected active channels of the hadron monitor. 
The typical accuracy of the beam-based alignment is $\pm0.1$\,mm, in agreement with optical survey measurements performed during the installation or replacement of beamline components.  

\begin{figure}[t]
\includegraphics[width=\linewidth]{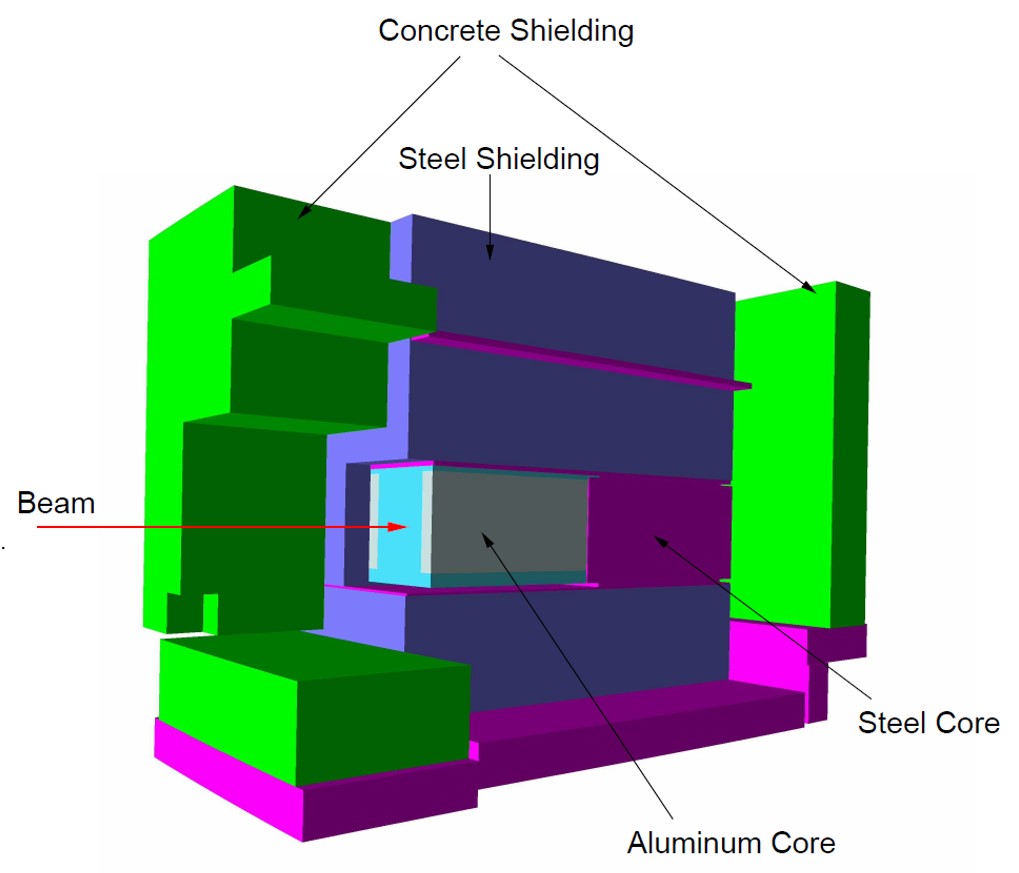}
\caption{\label{fig:abslayout}
The layout of the NuMI beam absorber.}
\end{figure}
\begin{figure}[t]
\includegraphics[width=1.0\linewidth]{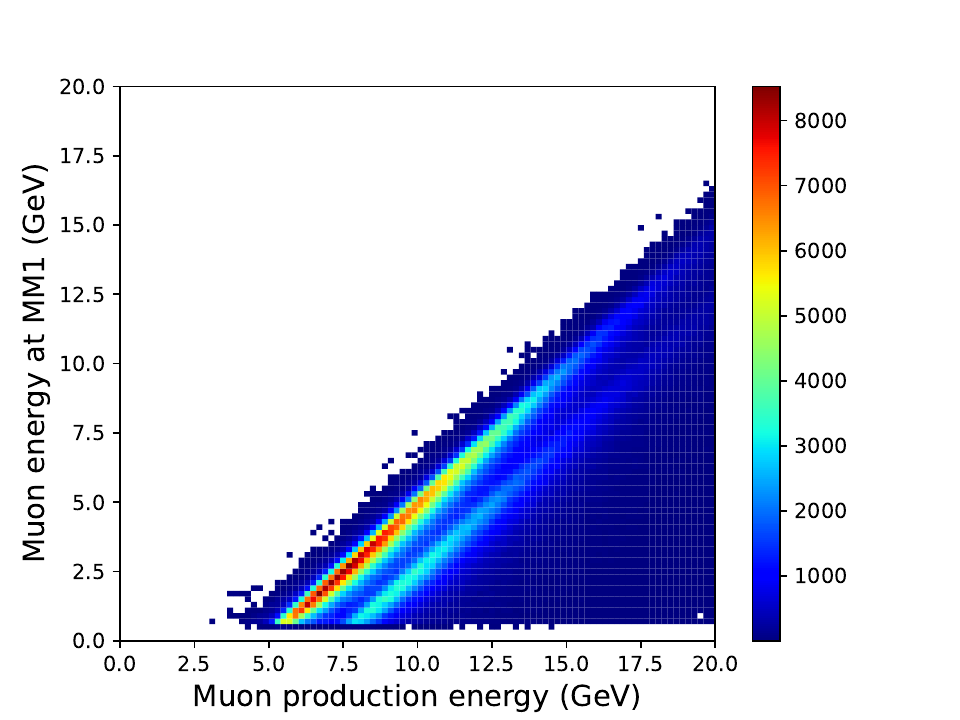}
\caption{\label{fig:abs}
Simulated 2D histogram of the muon spectrum. 
The horizontal axis is the muon energy at production, and the vertical axis is the muon energy at Muon Monitor~1.
The thick band is the muons passing through the aluminum core, while the second thin band is the muons passing through the steel block outside the aluminum core. 
Muons lose more energy in steel than in aluminum. 
}
\end{figure}

Figure~\ref{fig:abslayout} shows the hadron absorber---located immediately downstream of the hadron monitor---with some side shielding blocks removed for illustrative purposes. 
The absorber core consists of eight aluminum blocks at the upstream end, each measuring $1.3 \times 1.3 \times 0.3$\,m (width $\times$ height $\times$ length), followed by ten flame-cut steel blocks, each $1.3 \times 1.3 \times 0.23$\,m. 
Each aluminum block contains two independent water-cooling circuits, whereas the steel layers are not actively cooled. 
The steel blocks are designed to absorb the tails of the hadronic showers initiated in the aluminum core. 
Radiation shielding is provided by 88 additional steel blocks that surround the absorber core, effectively attenuating thermal and low-energy neutrons, to which the core is relatively transparent.
This double-layer absorber configuration is illustrated by the simulated muon flux distributions observed on the muon monitors (Fig.~\ref{fig:abs}). 
Finally, the entire assembly is enclosed in a concrete shield, which forms the hadron absorber and measures approximately $5.5 \times 5.6 \times 8.5$\,m.  

Three muon monitors are placed downstream of the absorber: 
Muon Monitor~1 (MM1), Monitor~2 (MM2), and Muon Monitor~3 (MM3), arranged sequentially from upstream to downstream. 
A detailed description of their construction and operational characteristics is provided in Sec.~\ref{subsec:muonmonitor}. 
Although most of the charged particles produced upstream are absorbed before reaching this region, muons penetrate the absorber and are subsequently detected by these monitors. 
All remaining particles, including muons, are stopped in the rock downstream of Muon Monitor~3. 
Only neutrinos continue to propagate beyond this point, reaching the neutrino detectors located further downstream.

\subsection{NuMI Beam Instrumentation}
\label{subsec:beaminstr}
This subsection describes the primary beam instrumentation relevant to muon profile studies, including muon monitors. 
\subsubsection{Proton Beam Monitor}
A pair of induced-charge pickup electrodes and a multiwire Secondary Emission electron Monitor (SEM) serve as the beam position monitor (BPM) and beam profile monitor (PM), respectively. 
A dedicated PM, constructed from thin titanium foil (\qty{5}{\micro\meter} thick $\times$ \qty{25}{\micro\meter} wide) to minimize beam scattering, is permanently installed 10\,m upstream of the target.  
This PM consists of $47 \times 47$ planar foils arranged in the $x$ and $y$ directions, with $0.5$\,mm spacing between adjacent foils. 
It records the proton beam profile for every beam spill. 
Additional PMs are periodically inserted at different locations along the beamline to support beam optics monitoring. 

Horizontal and vertical BPMs are paired to measure the beam position in the transverse plane. 
Two sets of paired BPMs, located $22$\,m and $10$\,m upstream of the target, are used to determine both the position and angular orientation of the proton beam at the target. 
These BPMs are part of an automated feedback system, known as the autotune~\cite{adamson2016}, which continuously adjusts the beam trajectory to maintain alignment with the target. 
The beam position can be measured with an accuracy of $\pm0.02$\,mm using BPMs and PMs~\cite{adamson2016}, and the RMS beam spot size can be determined with an accuracy of $\pm0.1$\,mm using PM. 

\subsubsection{Current Monitor}
The proton beam intensity at the target is measured by a toroidal beam current transformer (beam CT) installed upstream of the target. 
The CT signal is calibrated with a precision current source and cross-checked against a direct-current current transformer (DCCT) at the Main Injector. 
The beam CT achieves an accuracy of $0.5$\%, with a pulse-per-pulse stability of $\pm$0.1\%~\cite{adamson2016,Crisp_2011}. 
 
The horn power supply delivers a half-sinusoidal voltage waveform, which is transmitted through four striplines to the horns. 
The pulse duration is $2.334$\,msec. 
The horn current is measured using current transformers (horn CT) placed on each stripline. 
These devices are calibrated using a pre-calibrated current source, achieving a measurement accuracy of $\pm0.1$\%~\cite{Hylen2018}. 

\subsubsection{Muon Monitor}
\label{subsec:muonmonitor}
\begin{figure}[htbp]
\includegraphics[width=8.5cm]{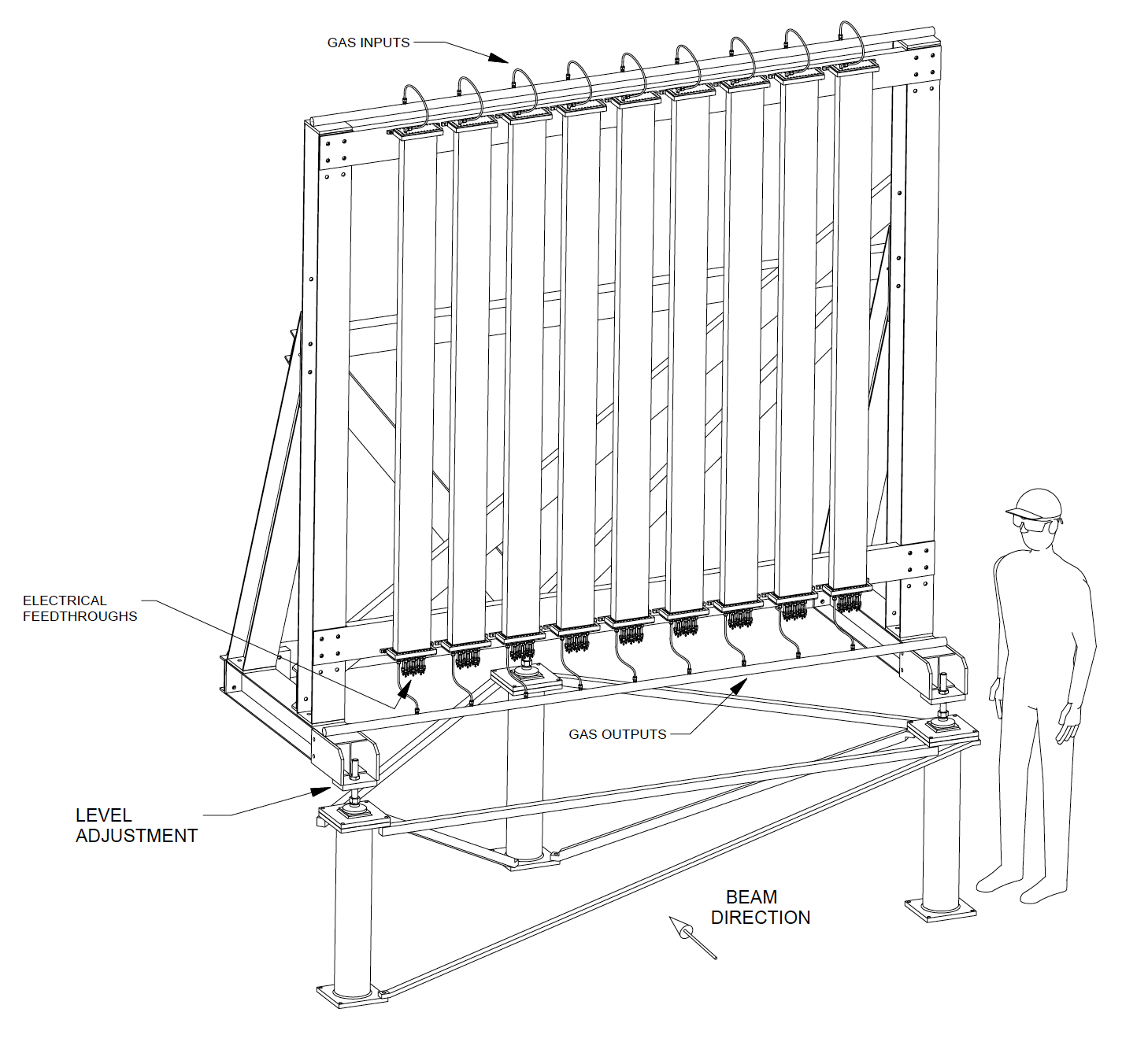}
\caption{\label{fig:mm_layout}
Engineering drawing of the muon monitor, showing the nine vertical tubes, each containing a row of nine ionization chambers.}
\end{figure}

Figure~\ref{fig:mm_layout} shows an engineering drawing of a muon monitor. 
Each monitor has a $9 \times 9$ grid of ionization chambers. 
Each chamber contains a $7.5 \times 7.5$\,cm$^2$ Ag-Pt plated electrode mounted on a 1-mm thick Alumina ceramic base, with a 3-mm gap between the electrodes. 
The channels are spaced $250$\,mm apart, covering a total active area of $2.1 \times 2.1$\,m$^2$. 
The nine ionization chambers are housed within a rectangular aluminum tube. 
Nine tubes are vertically assembled on a support structure. 
Each tube is filled with atmospheric pressure 99.995\% pure helium gas, which is injected from the top and exhausted from the bottom, serving as the ionization medium.  
A bias voltage of $300$\,V is applied, placing the chambers on the ionization plateau to ensure that all ionized particles are collected without charge multiplication~\cite{Loiacono2010,Zwaska2005}. 

\begin{figure}[tbp]
\includegraphics[width=7.5cm]{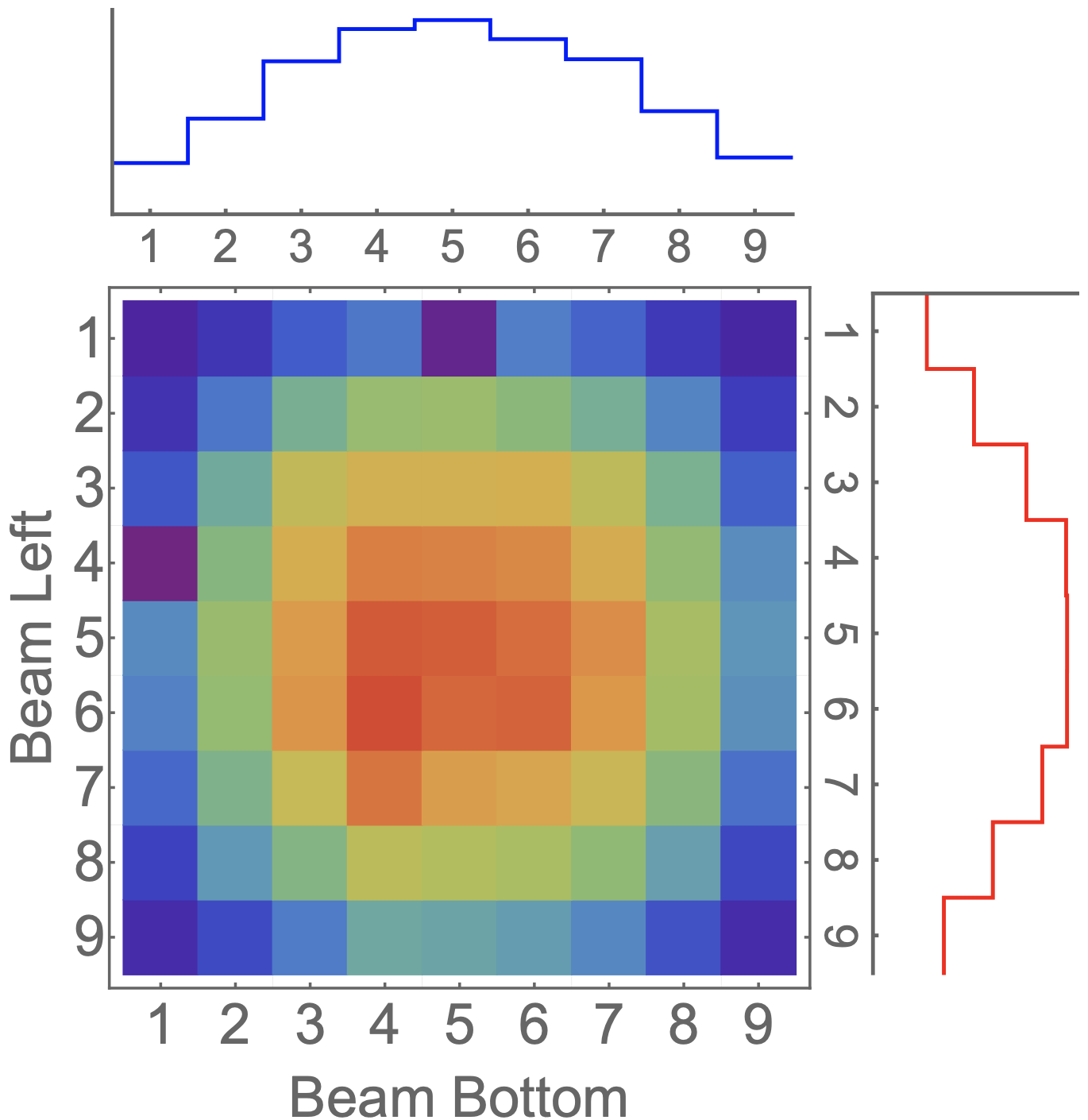}
\caption{\label{fig:mm1raw}
The observed beam profile at Muon Monitor~1. 
On Muon Monitor~1, there are two dead channels, (1,5) and (4,1). This analysis omits these dead channels.
The top and right subplots show the projection of the beam profile in horizontal and vertical planes, respectively. 
The scale of subplots is linear. 
}
\end{figure}

\begin{figure}
\includegraphics[width=7.5cm]{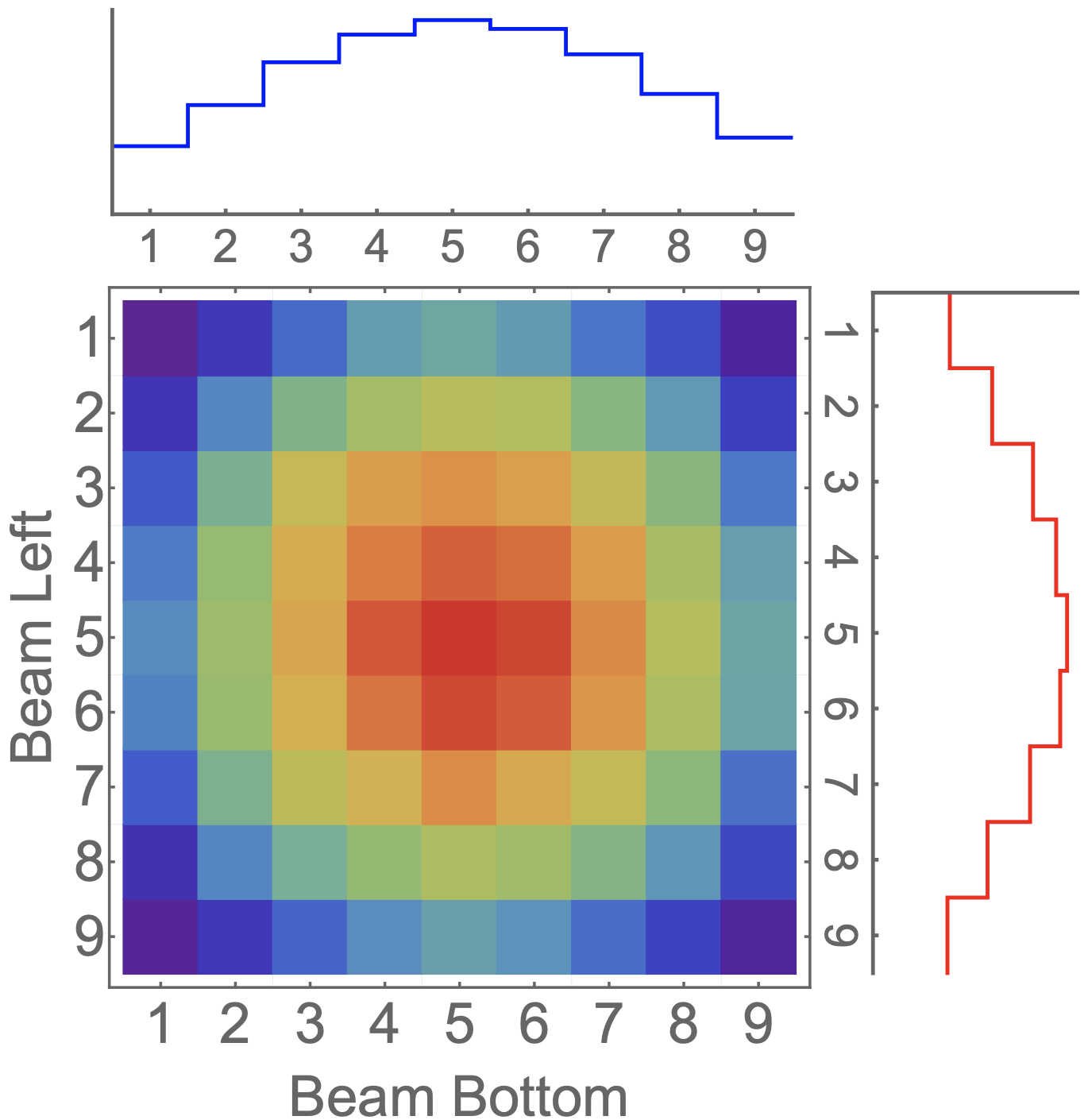}
\caption{\label{fig:mm2raw}
The observed beam profile at Muon Monitor~2.
}
\end{figure}

\begin{figure}
\includegraphics[width=7.5cm]{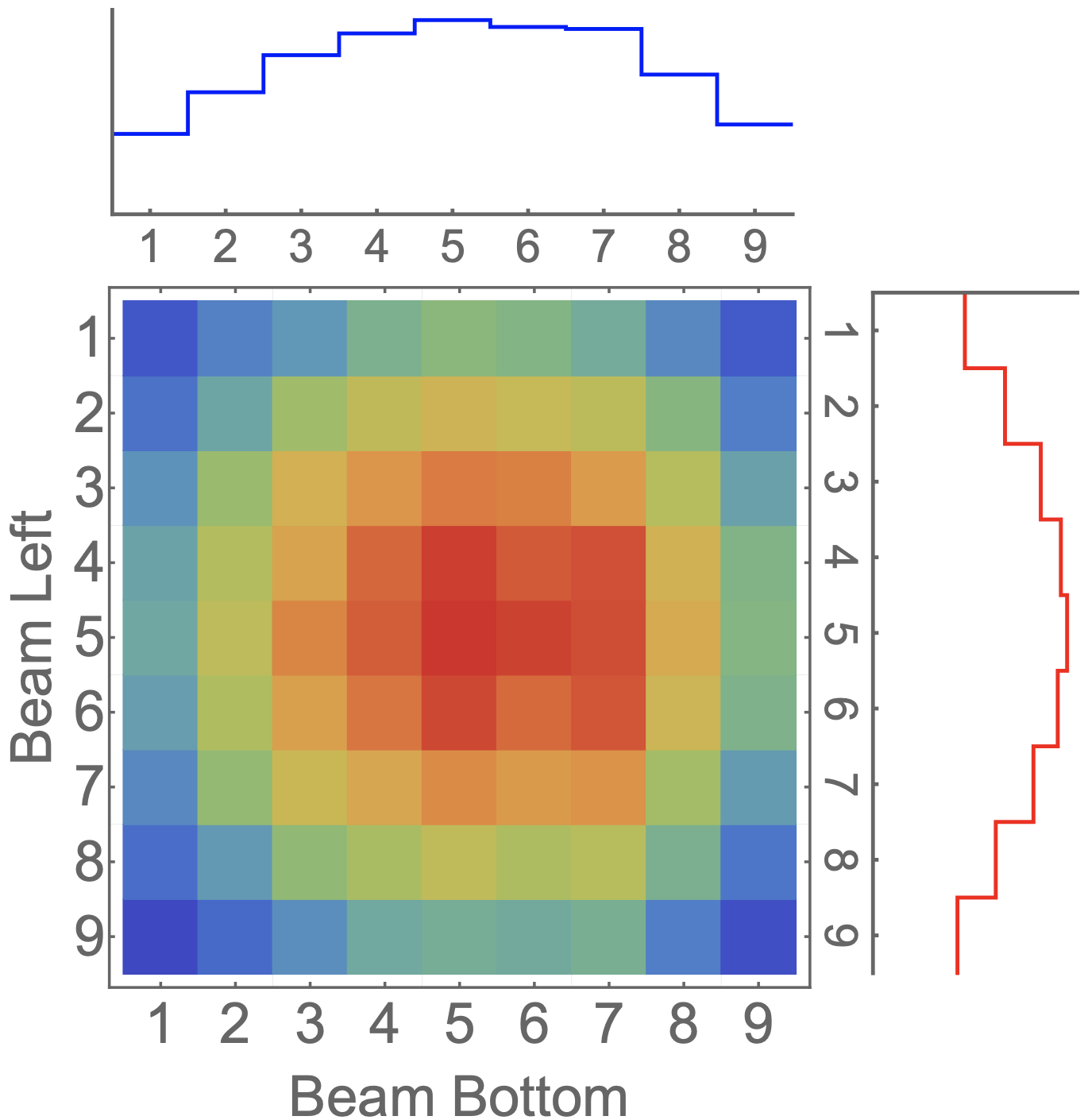}
\caption{\label{fig:mm3raw}
The observed beam profile at Muon Monitor~3.
}
\end{figure}

Figures~\ref{fig:mm1raw}--\ref{fig:mm3raw} display example pixel images recorded at each muon monitor during a single beam spill. 
The projections of the muon beam profile in the horizontal ($x$) and vertical ($y$) directions are shown with linear scaling. 
The pedestal signals, taken with the beam off, are subtracted from the beam-on signals. 
Pedestal levels are typically around $0.1$\% of the beam signal. 
Each channel was calibrated using a radiation source before beam exposure. 
During beam operation, the integrated signal across all channels is normalized to the proton beam intensity per spill and continuously monitored. 
The long-term fluctuation of the normalized signal lies within $\pm0.2$\% for Muon Monitors~1 and 2, and within $\pm0.5$\% for Muon Monitor~3. 
This stability confirms the linear response of the monitors to beam intensity. 
[tbp]
[tbp]

\begin{figure}[tbp]
\includegraphics[width=8.5cm]{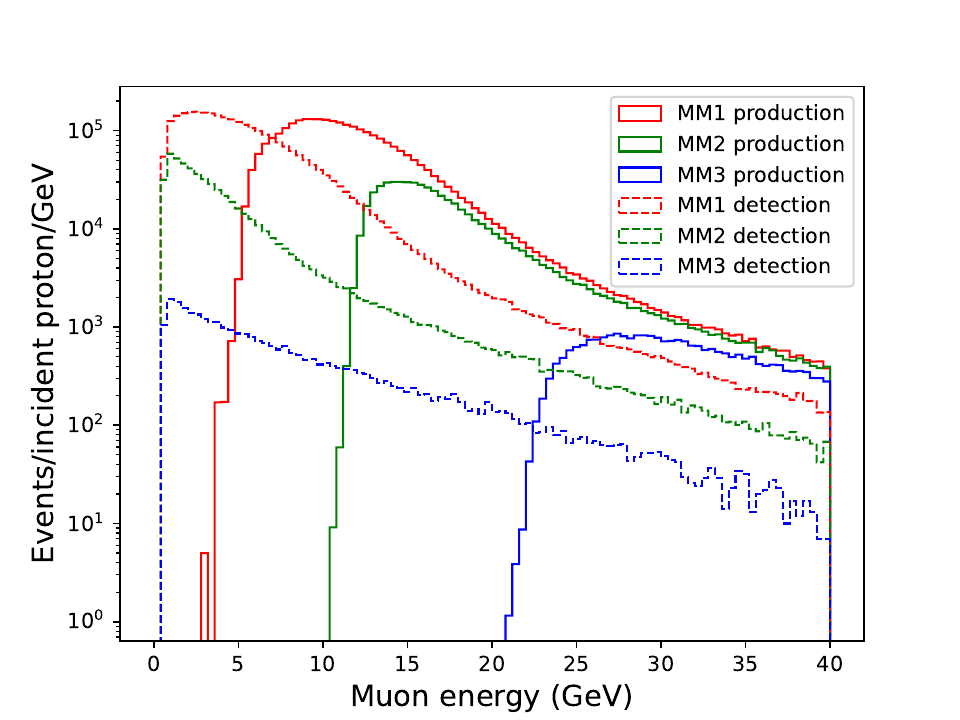}
\caption{\label{fig:mmspectra}
Simulated muon spectra. 
A solid red line represents the muon spectrum when muons are produced, while a dashed red line is the muon spectrum at Muon Monitor~1. 
The shift is because muons have lost energy in the absorber. 
The green and blue lines are for Muon Monitor~2 and~3, respectively. 
}
\end{figure}

\begin{figure}[btp]
\includegraphics[width=8.5cm]{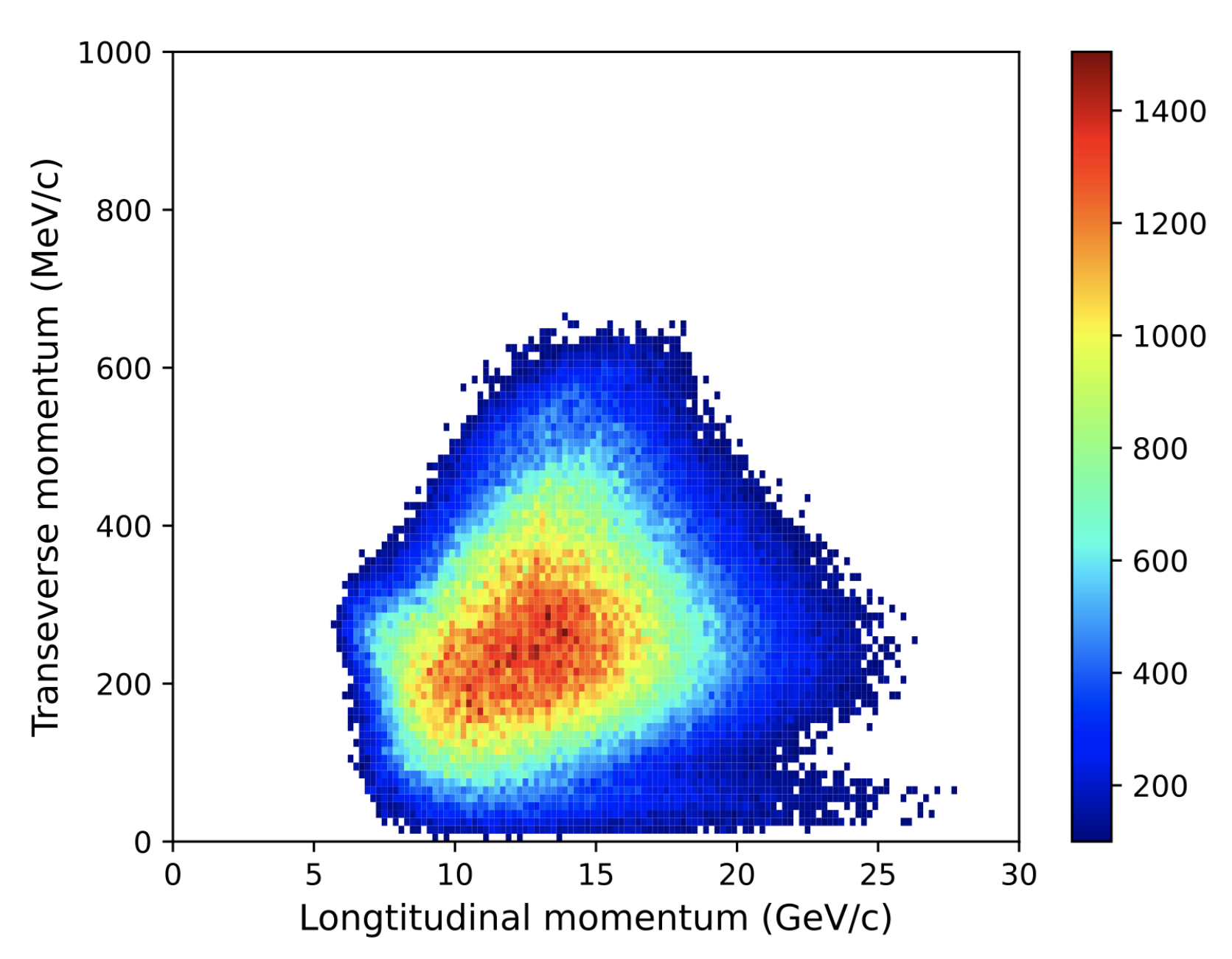}
\caption{Simulated initial transverse and longitudinal momenta of the pions for which the corresponding muons reach Muon Monitor~1. It should be noted that the low statistic bins are cut and set to zero in the plot.}
\label{fig:mm1pion}
\end{figure}

\begin{figure}[btp]
\includegraphics[width=8.5cm]{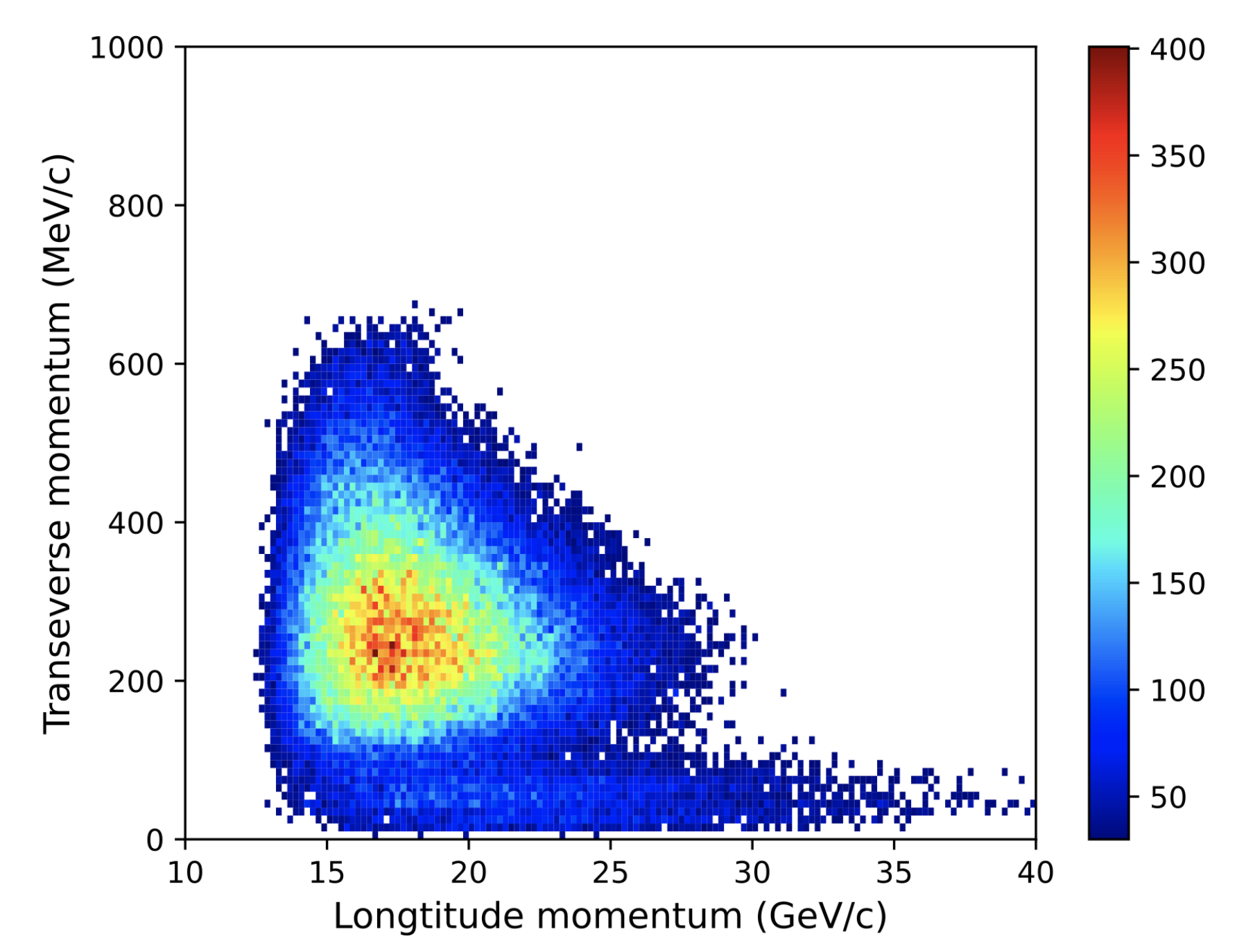}
\caption{Simulated initial transverse and longitudinal momenta of the pions for which the corresponding muons reach Muon Monitor~2.}
\label{fig:mm2pion}
\end{figure}

\begin{figure}[btp]
\includegraphics[width=8.5cm]{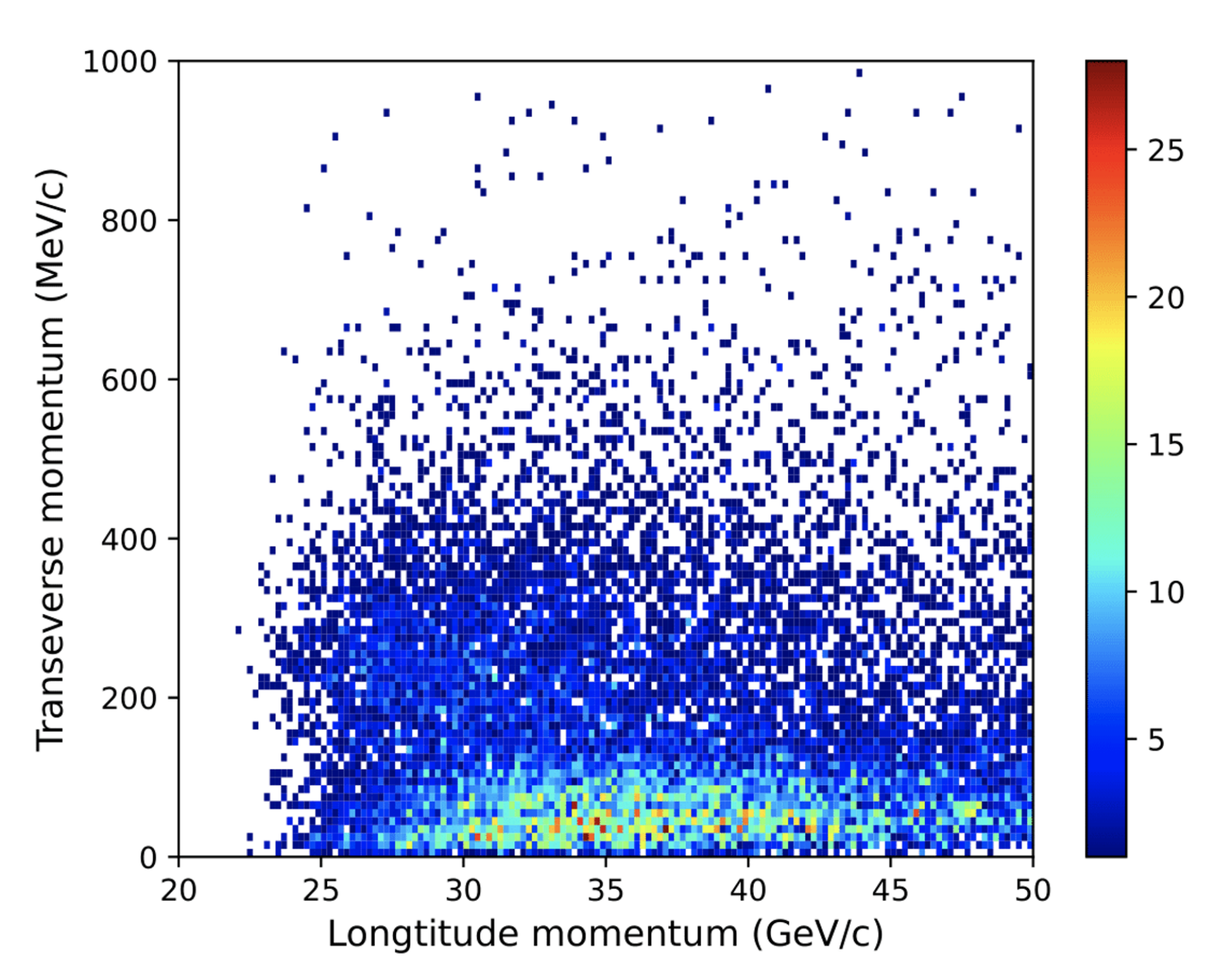}
\caption{Simulated initial transverse and longitudinal momenta of the pions for which the corresponding muons reach Muon Monitor~3.}
\label{fig:mm3pion}
\end{figure}

Figure~\ref{fig:mmspectra} shows that each monitor detects a unique muon energy spectrum, with approximate cutoff energies of $7$\,GeV, $12$\,GeV, and $23$\,GeV for MM1, MM2, and MM3, respectively. 
Figures~\ref{fig:mm1pion}--\ref{fig:mm3pion} show the simulated initial momentum phase space of the pions whose daughter muons reach each corresponding muon monitor. 
These pion momentum ranges are consistent with the simulated muon spectra observed at each monitor. 
Multiple Coulomb scattering affects muon trajectories as they pass through materials, resulting in complex signals that cannot be directly linked to specific neutrino events detected in the detectors.
Therefore, muon signals are not tagged with individual neutrino interactions in the detectors.

\section{Horn focusing mechanism}
\label{sec:hornfocusingmechanism}
Modern computational tools enable the design and optimization of advanced horn focusing systems, including tuning proton beam parameters, target geometry, and the complex structure of multiple horn configurations. 
However, this optimization introduces significant complexity, complicating the focusing mechanism analysis. 
We adopt the following multi-step approach to study the horn focusing mechanism systematically. 
First, we introduce an analytical model that captures the fundamental focusing properties of a single magnetic horn. 
We then extend this to a semi-analytical model that accounts for the full NuMI multi-horn system, enabling a more comprehensive study of its focusing behavior. 
Finally, we perform numerical particle tracking simulations to validate the models and assess the linearity of the focusing response in the realistic horn geometry by comparing results with analytical predictions. 

\subsection{Analytical Model of Single Horn Focusing}
\begin{figure}
\includegraphics[width=\linewidth]{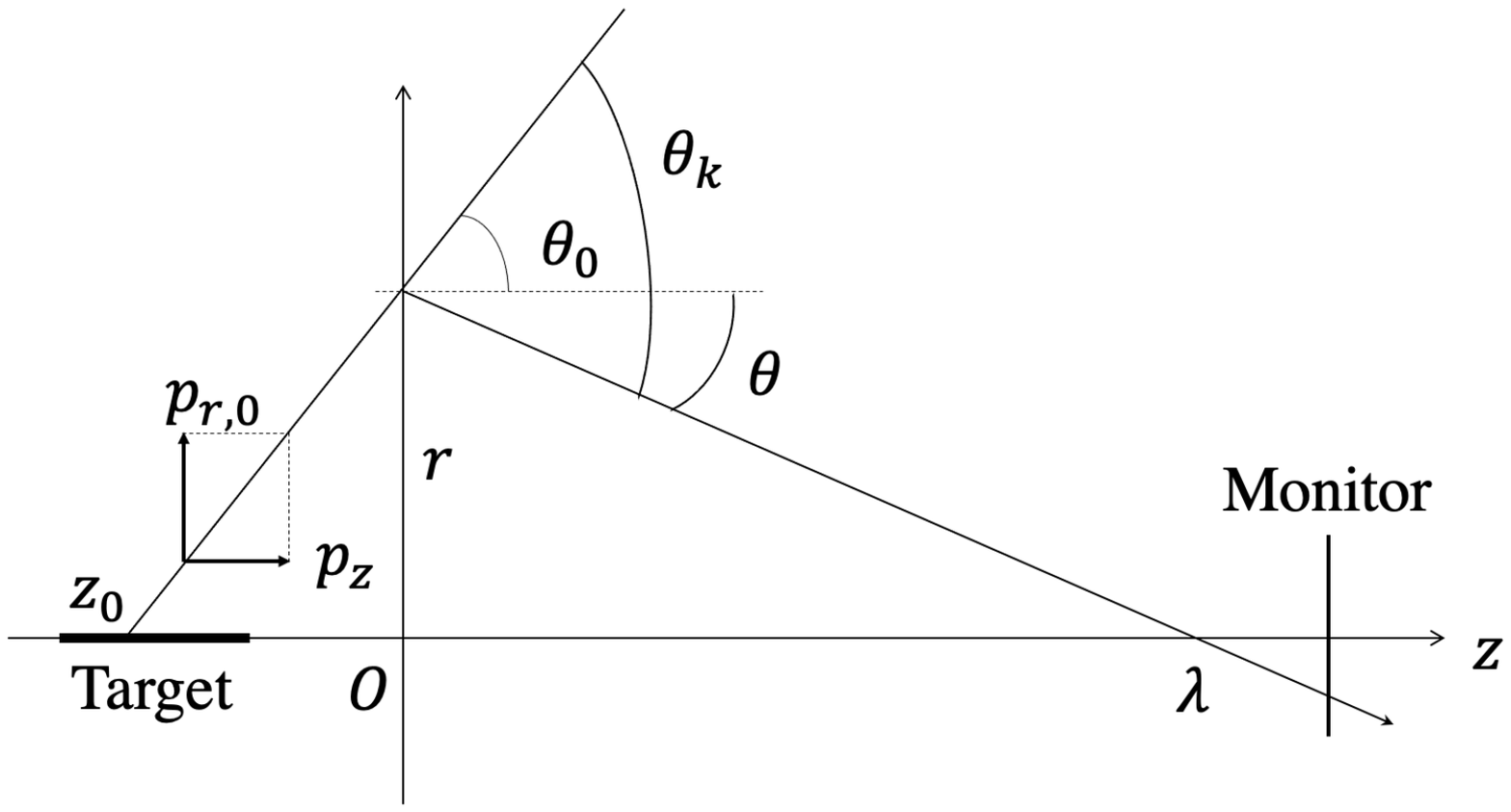}
\caption{\label{fig:raytrace_diagram}
Schematic of the geometry used in the analytical model. 
Point ``$\it{O}$'' denotes the horn's position. 
In this model, the horn is treated as having zero longitudinal extent, and the transverse kick is applied instantaneously, as described by Eq.~(\ref{eq:anamodel}). 
The target has a finite length, and pions are assumed to originate at longitudinal position $z=z_0$ with initial momenta $p_{r,0}$ and $p_z$. 
A monitor is placed $z = 700$\,m downstream along the beam axis. 
Because the distances between the three monitors are small compared to the length of the decay pipe, just one monitor is used in the analytical model. 
}
\end{figure}

The analytical expression for the transverse momentum kick imparted to a charged particle by a single magnetic horn is given in Eq.~(\ref{eq:simple3}) in Appendix~\ref{sec:appendix_analytical}. 
The resulting deflection angle, $\theta$, is given by:
\begin{equation}
\theta \equiv \theta_0-\theta_k = \theta_0-\frac{\mu_0 I q}{2 \pi}\frac{ar}{p_z},
\label{eq:anamodel}
\end{equation}
where $\theta_0$ and $\theta$ are the particle angles before and after passing through the horn, respectively, $\theta_k$ the transverse kick imparted by the horn, $\mu_0$ the vacuum permeability, $I$ the horn current, $q$ the particle charge, $p_z$ the longitudinal momentum, $a$ the parabolic profile coefficient of the inner conductor, and $r$ the radial position of the incident particle at the upstream face of the horn. The geometrical configuration used in the analytical model is illustrated in Fig.~\ref{fig:raytrace_diagram}. 

We define a focusing parameter, $\lambda$, as the axial distance from the horn to the point along the $z$ axis where the particle’s trajectory intersects the beam axis,
\begin{equation}
\lambda = \frac{r}{\tan \theta} \sim \frac{r}{\theta} = \frac{r}{\theta_0 - \frac{\mu_0 q I}{2\pi} \frac{a r}{p_z}}.
\label{eq:analyticalmodel}
\end{equation}
\begin{figure}
\includegraphics[width=\linewidth]{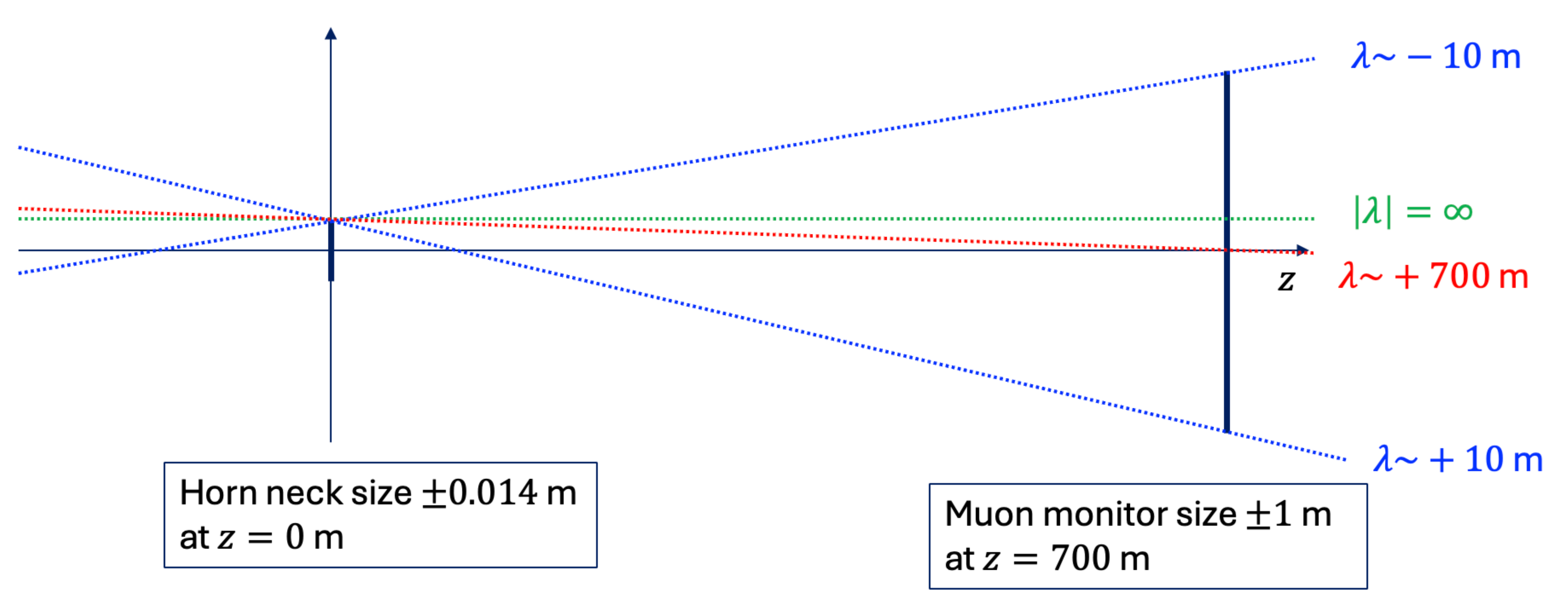}
\caption{\label{fig:lambda}
The focusing parameter $\lambda$. 
}
\end{figure}
Figure~\ref{fig:lambda} shows the geometric definition of $\lambda$. 
Based on the value of $\lambda$, we classify the focusing behavior into five distinct regimes:  
\begin{itemize}
\item A focusing length is identified when $|\lambda|\rightarrow\infty$. 
This occurs when the transverse kick cancels the initial angle $\theta_0$, as described by Eq.~(\ref{eq:focuscondition}). 
\item When $10$\,m $< \lambda < 700$\,m, the particle trajectory crosses the beam axis upstream of the muon monitor. 
This condition is referred to as "overfocus". 
\item When $700$\,m $< \lambda < +\infty$, the trajectory intersects the beam axis downstream of the monitor. 
This condition is referred to as "underfocus". 
\item When $-\infty < \lambda < -10$\,m, the trajectory intersects the beam axis upstream of the horn yet the particle still passes through the monitor. 
This case is referred to as "divergent". 
\item When $\lambda\sim 700$\,m, the trajectory intersects the beam axis near the location of the monitor. 
This condition is referred to as "in focus at the muon monitor". 
\end{itemize}
The particle trajectory does not cross the muon monitor when $-10$\,m $< \lambda < 10$\,m.

\subsection{Semi-Analytical Model of Multi-Horn Focusing}
\label{sec:semi-analytical_model}
The NuMI beamline uses two magnetic horns, each with a parabolic shape at both upstream and downstream ends, as illustrated in Fig.~\ref{fig:horncurrent}. 
This geometry effectively creates two focusing sections within each horn. 
However, the analytical model applies only a single-horn configuration and therefore cannot fully represent the NuMI horn system. 
To simulate the NuMI horn configuration more accurately, we introduce a semi-analytical multi-horn model. 
In this model, the transverse kick is calculated using the analytical expression given in Eq.~(\ref{eq:anamodel}).
When a particle travels between Horn~1 and Horn~2, the particle is assumed to travel along a straight-line trajectory in vacuum. 
This hybrid approach---combining transverse kicks within the horns and straight-line propagation in field-free regions---is referred to as the ``semi-analytical model". 
The horn geometry used in the model---specifically, the dimensions of the inner and outer conductors for Horn~1 and Horn~2---is summarized in Table~\ref{tab:horngeometry}. 
Because the neck of Horn~1 is located $z = 0.8$\,m downstream from the origin, we introduce a $-0.005$\,m offset to the initial radius $r$ in the semi-analytical model to account for this displacement. 
\begin{table}[b]
\caption{\label{tab:horngeometry}%
Geometric profiles of the inner and outer conductors of Horn~1 and Horn~2, as used in the semi-analytical model.  
Here, $z$ denotes the longitudinal position along each horn.
}
\begin{ruledtabular}
\begin{tabular}{ccc}
 & $z$ (mm) & profile (mm) \\
 \hline
 Horn 1 inner & $0.0 \leq z < 800.0$ & $\displaystyle{\sqrt{\frac{928.5 - z}{70.5}}}$ \\
              & $800.0 \leq z < 839.8$ (neck) & $13.5$ \\[2pt]
              & $839.8 \leq z < 3000.0$ & $\displaystyle{\sqrt{\frac{z - 800.0}{21.9}}}$ \\
 \hline
 Horn 1 outer & $0.0 \leq z < 3000.0$ & $149.2$ \\
 \hline
 Horn 2 inner & $0.0 \leq z < 976.2$ & $\displaystyle{\sqrt{\frac{1000.0 - z}{1.4}}}$ \\
              & $976.2 \leq z < 1048.0$ (neck) & $39.0$ \\[2pt]
              & $1048.0 \leq z < 3000.0$ & $\displaystyle{\sqrt{\frac{z - 1000.0}{2.7}}}$ \\
 \hline
 Horn 2 outer & $0.0 \leq z < 3000.0$ & $370.0$ \\
\end{tabular}
\end{ruledtabular}
\end{table}

\subsection{Numerical Validation of Analytical Models}
We use G4beamline~\cite{g4beamline} to perform particle tracking simulations that evaluate the analytical and semi-analytical models of the NuMI horn focusing mechanism.
The simulation includes actual geometric components such as the horn conductors, beam windows, and ambient air. 
A magnetic field is present between the inner and outer conductors of the horns, and the region is filled with atmospheric pressure argon gas. 
The simulation includes the Lorentz force in the magnetic fields and electromagnetic energy loss in materials.
However, stochastic processes such as pion decay, multiple scattering, and hadronic interactions are disabled. 
Test particles (pions) are initially placed at $z= 0$\,m and uniformly distributed in radius with a step size of $0.5$\,mm. 
The initial momentum is varied depending on the specific study case. 
The particle parameters $(r, z, p_r, p_z)$ are recorded downstream of Horn~2 to calculate the focusing parameter $\lambda$. 

\begin{figure}
\includegraphics[height=\linewidth,angle=-90]{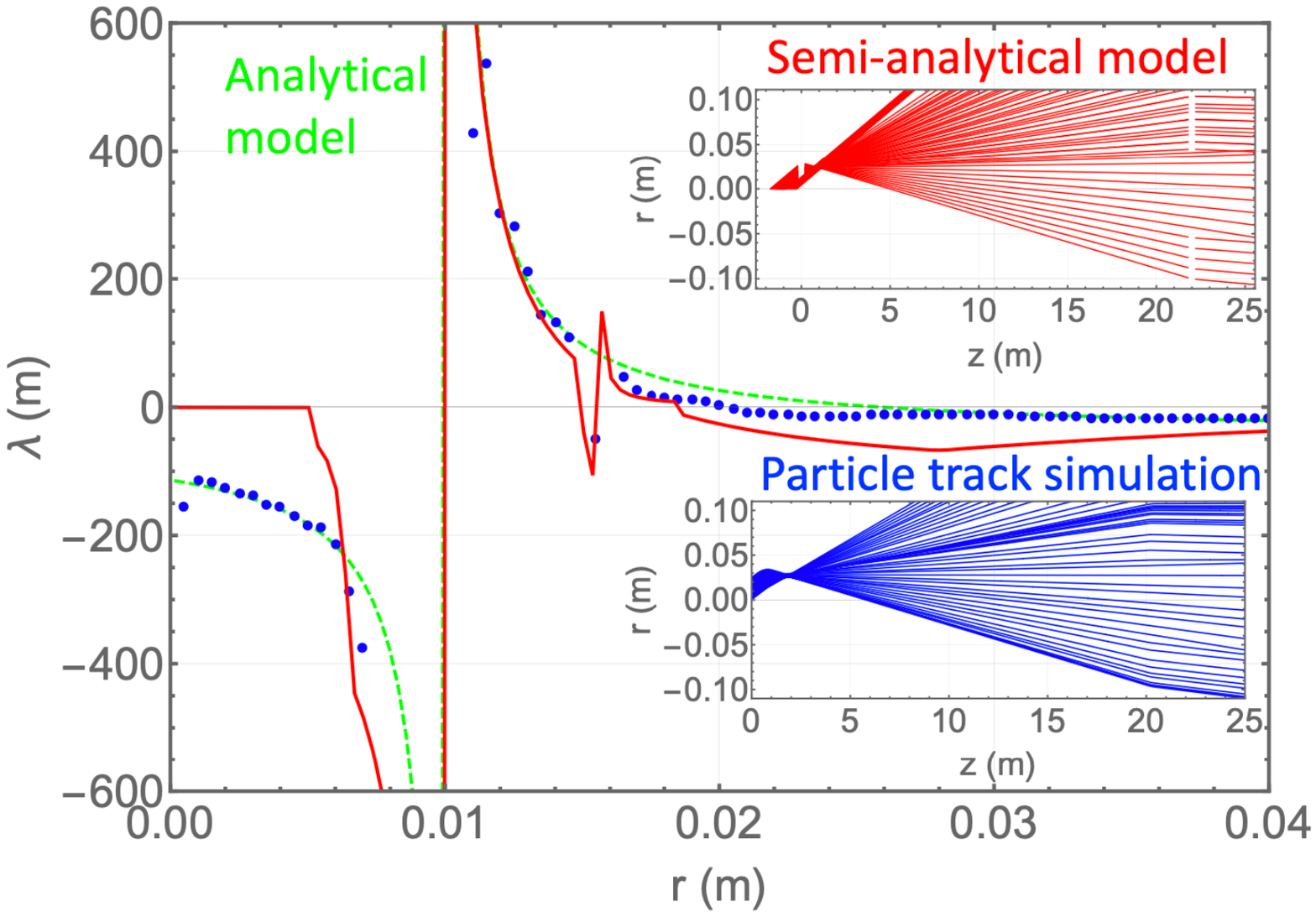}
\caption{Simulated $\lambda$ as a function of the initial radial position $r$ for a pion with $p_z = 12$\,GeV/$c$ and $p_{r,0} = 0.2$\, GeV/$c$, representative of muons reaching Muon Monitor~1. 
Blue points: G4beamline simulation; solid red line: semi-analytical model; dashed green line: fitted analytical model.
Top subplot: semi-analytical model ray trace; bottom subplot: simulation ray trace.}
\label{fig:anavalid1}
\end{figure}

Figure~\ref{fig:anavalid1} shows the estimated $\lambda$ as a function of the initial radial position $r$.
The results are presented for a pion with $p_z = 12$\,GeV/$c$ and $p_{r,0} = 0.2$\,GeV/$c$. 
This specific momentum configuration is representative of the initial pion momenta that contribute predominantly to the production of muons reaching Muon Monitor~1 (see Fig.~\ref{fig:mm1pion}).
Blue points represent the simulation results, the solid red curve corresponds to the semi-analytical model, and the dashed green curve shows the analytical model. 
Two singularities occur where $|\lambda| \rightarrow \infty$ at $r = 0.01$\,m and $r = 0.015$\,m. 
The semi-analytical model agrees well with the simulation result for $r > 0.01$\,m including two singularities. 
The ray traces from both the semi-analytical model and the simulation (shown in the subplots of Fig.~\ref{fig:anavalid1}) indicate that the second singularity at $r = 0.015$\,m occurs when the pions experience a kick in Horn~1, cross the beam axis, and are then kicked again by Horn~2, resulting in trajectories that are parallel to the beam axis. 
However, the semi-analytical model fails to reproduce the simulated $\lambda$ values for $r < 0.01$\,m. 
This discrepancy likely arises because, in the analytical model, the horn has zero longitudinal extent, causing particles with $r < 0.01$\,m to miss the horn and experience no kick. 

The green dashed line represents a fitted analytical model, where the parabolic profile coefficient $a$ in Eq.~(\ref{eq:analyticalmodel}) has been adjusted to best match the simulation results. 
The best fit is obtained with $a = 69$\,mm$^{-1}$, where the analytical curve closely matches most simulation points, except for the singularities. 
This result suggests that, despite simplifications, the single horn model reasonably captures the effective focusing behavior of the horn system. 
The fitted coefficient $a$ does not match any individual values of the inner conductor shape listed in Table~\ref{tab:horngeometry}; instead, it appears to represent an effective average, analogous to the combined focusing strength of multiple lenses. 

\begin{figure}
\includegraphics[height=\linewidth,angle=-90]{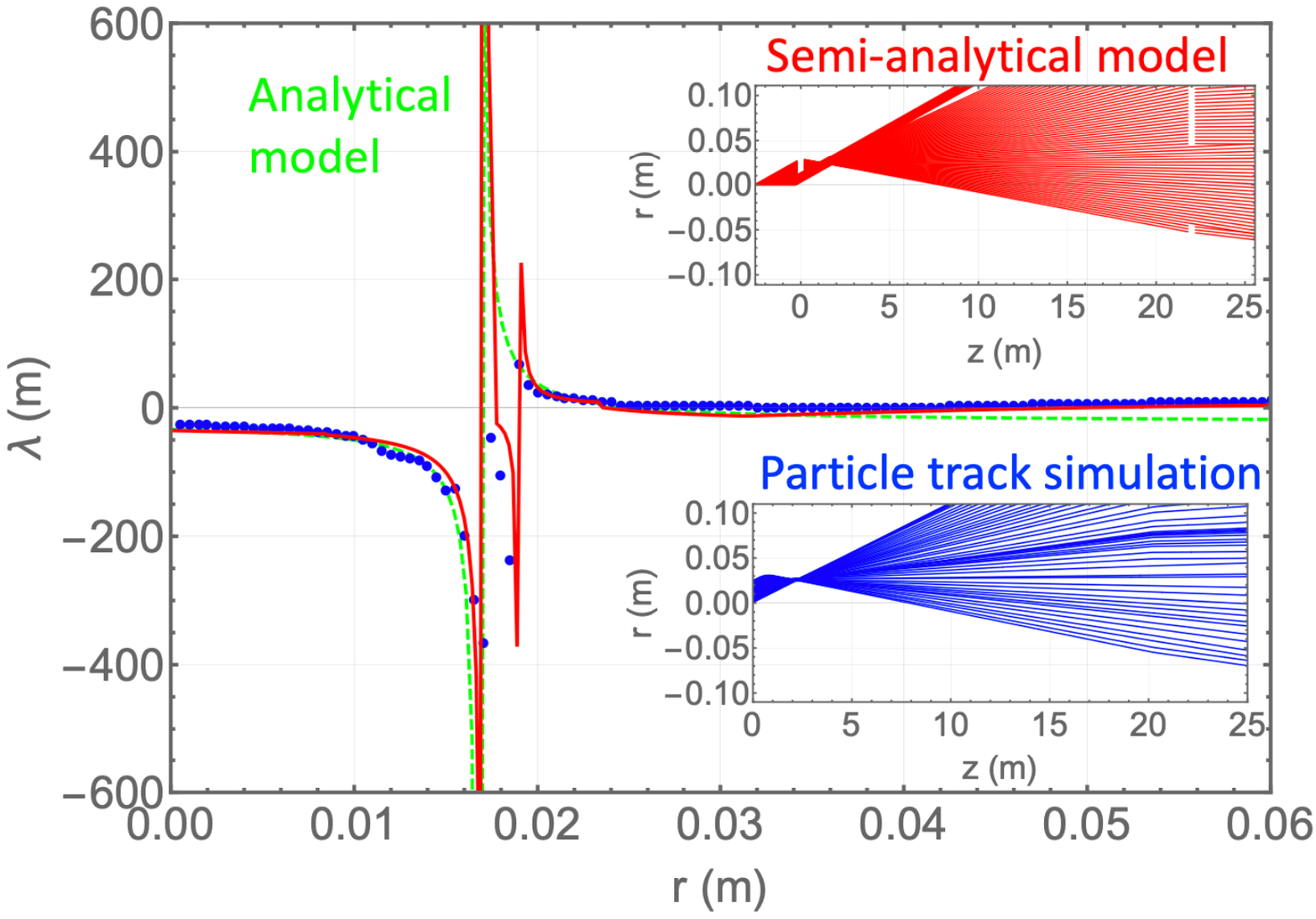}
\caption{Simulated $\lambda$ as a function of the initial radial position $r$ for a pion with $p_z = 18$\,GeV/$c$ and $p_{r,0} = 0.2$\,GeV/$c$, representative of muons reaching Muon Monitor~2. 
Blue points: G4beamline simulation; solid red line: semi-analytical model; dashed green line: analytical model.
Top subplot: semi-analytical model ray trace; bottom subplot: simulation ray trace. 
}
\label{fig:anavalid2}
\end{figure}

Figure~\ref{fig:anavalid2} shows the simulated $\lambda$ for pions with $p_z = 18$\,GeV/$c$ and $p_{r,0} = 0.2$\,GeV/$c$. 
This specific momentum configuration is representative of the initial pion momenta that contribute predominantly to the production of muons reaching Muon Monitor~2 (see Fig.~\ref{fig:mm2pion}).
The semi-analytical model accurately reproduces both singularities observed in the simulation. 
Furthermore, the simulation results align well with the analytical model when the parabola coefficient is set to $a = 48$\,mm$^{-1}$. 
The semi-analytical model also reproduces the simulation results for $r < 0.017$\,m. 
These pions intersect Horn~2, although this region lies outside the plotting range in the subplot. 

\subsection{Evaluation of NuMI Horn Acceptance with Full Simulation and Semi-Analytical Model}
The acceptance of the NuMI horn system is evaluated using both a full model numerical simulation and the semi-analytical model.
The full simulation is based on G4NuMI, a Geant4-based~\cite{geant4} particle tracking framework that includes the most up-to-date geometry and components of the NuMI beamline. 
This simulation is typically used for neutrino flux predictions. 
For this study, we customized G4NuMI to enable detailed muon tracking and enhanced event statistics by duplicating pion events multiple times. 

The full model simulation starts with a proton beam striking the target. 
All resulting particles---primary, secondary and tertiary---are tracked along the NuMI beamline, excluding electrons, neutrons, and gamma rays.  
The simulation accounts for particle interactions with all relevant materials, including the target, beam windows, horn structures including cooling water, decay pipe, and intervening gases and air. 
It also includes the hadron absorber and surrounding rocks up to the muon monitors. 
Although particles produced in the absorber and surrounding rock are not subject to the horn focusing mechanism, they are included in the simulation and contribute to the muon monitor signal. 
All relevant stochastic processes---such as decays, scattering, and hadronic interactions---are enabled to accurately model particle transport.  
In this framework, the acceptance is defined as the initial momentum phase space of pions---evaluated immediately after exiting the relevant materials---whose decay products (muons) reach the muon monitors. 
In contrast, the semi-analytical model does not simulate the proton-target interaction or any stochastic processes. 
Instead, it begins with an ensemble of pions uniformly distributed along the $z$ positions corresponding to the target surface, with both transverse and longitudinal momenta sampled uniformly. 
As a result, the semi-analytical model yields only the geometric acceptance boundary, rather than a probability density function. 

\begin{figure}
\includegraphics[width=\linewidth]{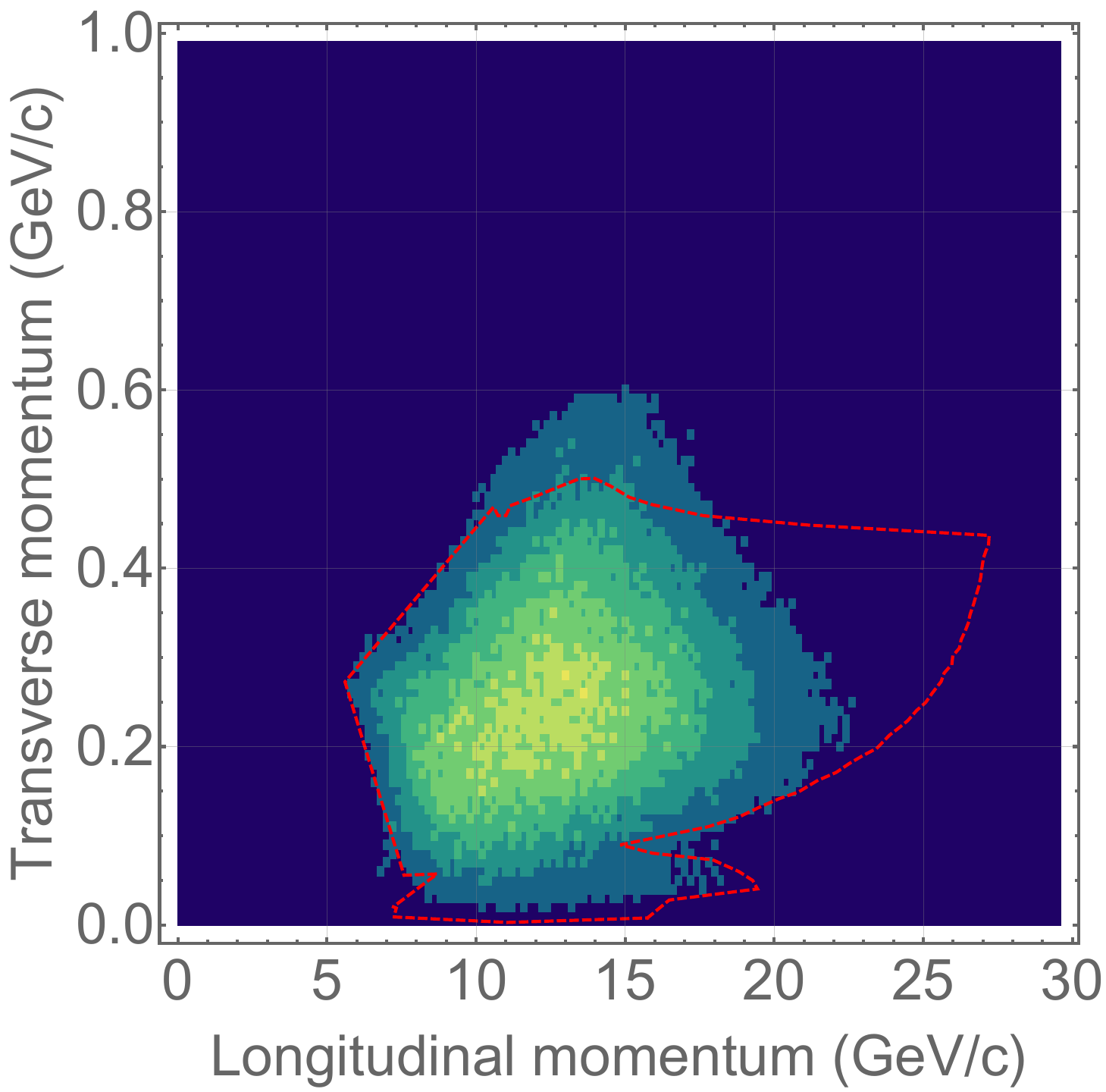}
\caption{Full model simulation result (2D histogram) vs. semi-analytical model (red dashed boundary outline).}
\label{fig:accept1}
\end{figure}

Figure~\ref{fig:accept1} compares the full model simulation and the semi-analytical model. 
In the full simulation, the acceptance is shown as a 2D histogram, where the color scale represents the probability density function. 
The red dashed line indicates the acceptance boundary calculated from the semi-analytical model. 
This distribution reflects the likelihood that pions with given initial conditions will produce muons that reach the monitors. 
The two approaches show good overall agreement, validating the prediction of the acceptance boundary from the semi-analytical model. 
Discrepancies are observed in the high $p_z$ and high $p_{r,0}$ regions. 
Most pions with $p_z \ge 25$\,GeV/$c$ do not decay before reaching the hadron absorber. 
These pions are absorbed in the hadron absorber, and a small fraction produces high-energy muons in the full model simulation. 
The semi-analytical model does not account for this pion loss mechanism, resulting in an overestimation of acceptance in this region. 
Additionally, pions with $p_{r,0} \ge 0.5$\,GeV/$c$ miss the horns in the semi-analytical model because of the assumption of zero horn length. 
This simplification contributes to the underestimation of acceptance in this region. 

\begin{figure}[tb]
\includegraphics[width=6.5cm,angle=-90]{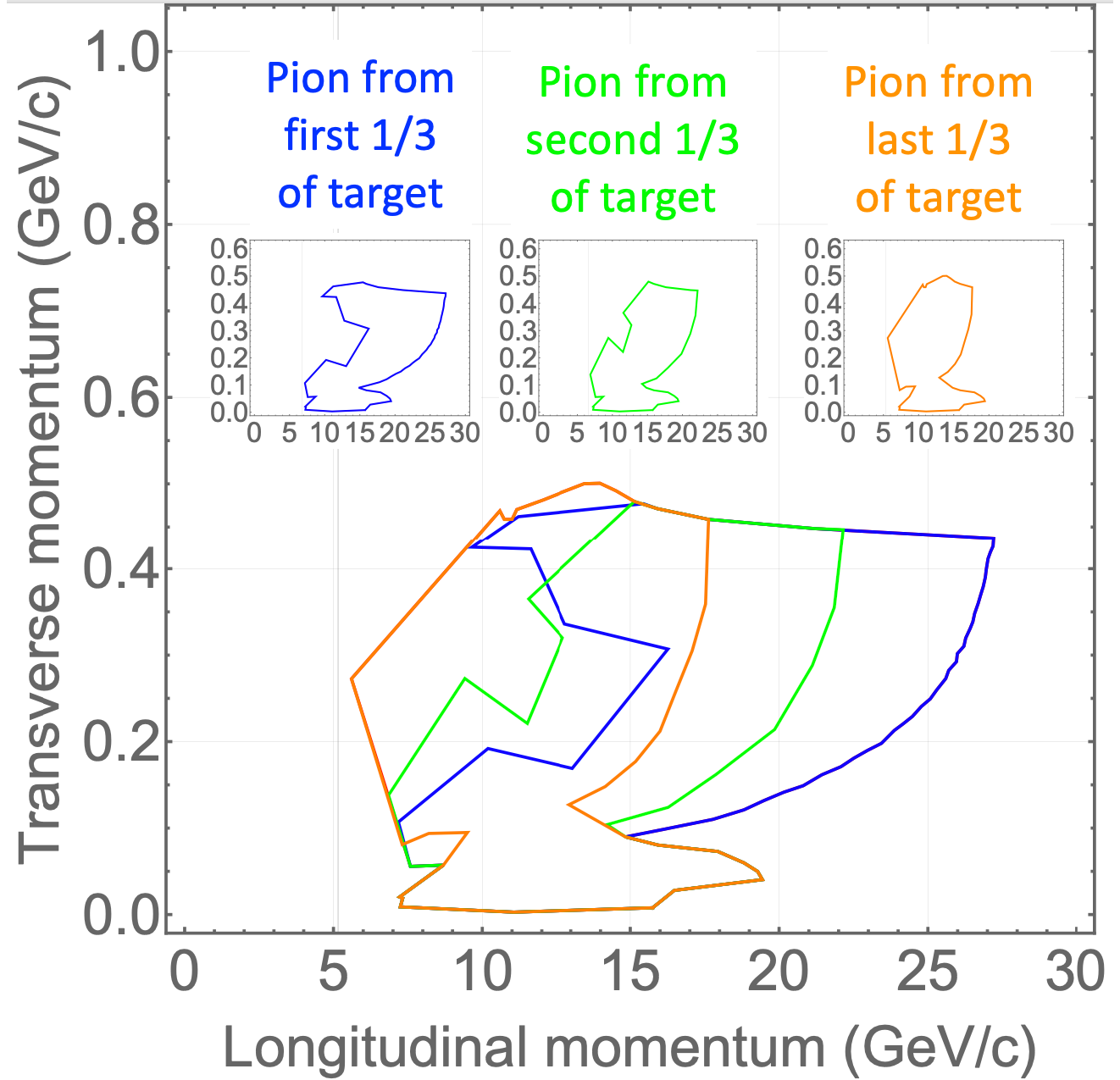}
\caption{
Simulated acceptance boundaries from the semi-analytical model, color-coded by the pion emission position along the target: blue (upstream third), green (middle third), and orange (downstream third). 
}
\label{fig:accept2}
\end{figure}

To further demonstrate the utility of the semi-analytical model, we apply it to investigate how acceptance varies with pion emission position. 
Figure~\ref{fig:accept2} demonstrates how the acceptance depends on the pion emission position along the target length. 
The blue, green, and orange regions correspond to pions originating from the upstream, middle, and downstream thirds of the target, respectively. 
Higher-momentum pions emitted from upstream third are more likely to reach the monitors. 
Conversely, lower-momentum pions from the downstream third are more likely to be focused to the monitors. 
It suggests a spatial-energy correlation in horn focusing: Muon Monitor~2 and 3, sensitive to high-energy muons, are most responsive to pion production in the upstream third of the target. Conversely, Muon Monitor~1 is more influenced by pions emitted from the downstream region.

\section{Beam Response Measurement}
\label{sec:beamresponsemeasurement}
The first systematic measurement of the NuMI horn focusing response was carried out on December 12, 2019. 
This measurement involved varying both the position of the proton beam on the target and the horn current. 
The procedure and beam conditions are summarized below:
\begin{itemize}
\item The horn operated in FHC mode. 
\item The proton beam intensity was limited to less than $4 \times 10^{13}$ POT per spill to protect the target from potential damage due to off-center beam incidence. 
\item The proton beam position at the target was scanned in both horizontal and vertical directions while maintaining the beam orientation parallel to the horn axis. 
The center of the target, as determined from the proton tomography, was located at (horizontal, vertical) = ($0.0$, $-0.8$)\,mm. 
The beam position was varied within approximately $\pm0.8$\,mm from this center. 
\item The muon beam profile at each muon monitor was recorded as a function of the position of the proton beam at the target.
\item After completing the beam position scan, the horn current was varied among $200$, $195$, $190$, and $180$\,kA. 
\end{itemize}

\begin{figure}
\includegraphics[width=8cm]{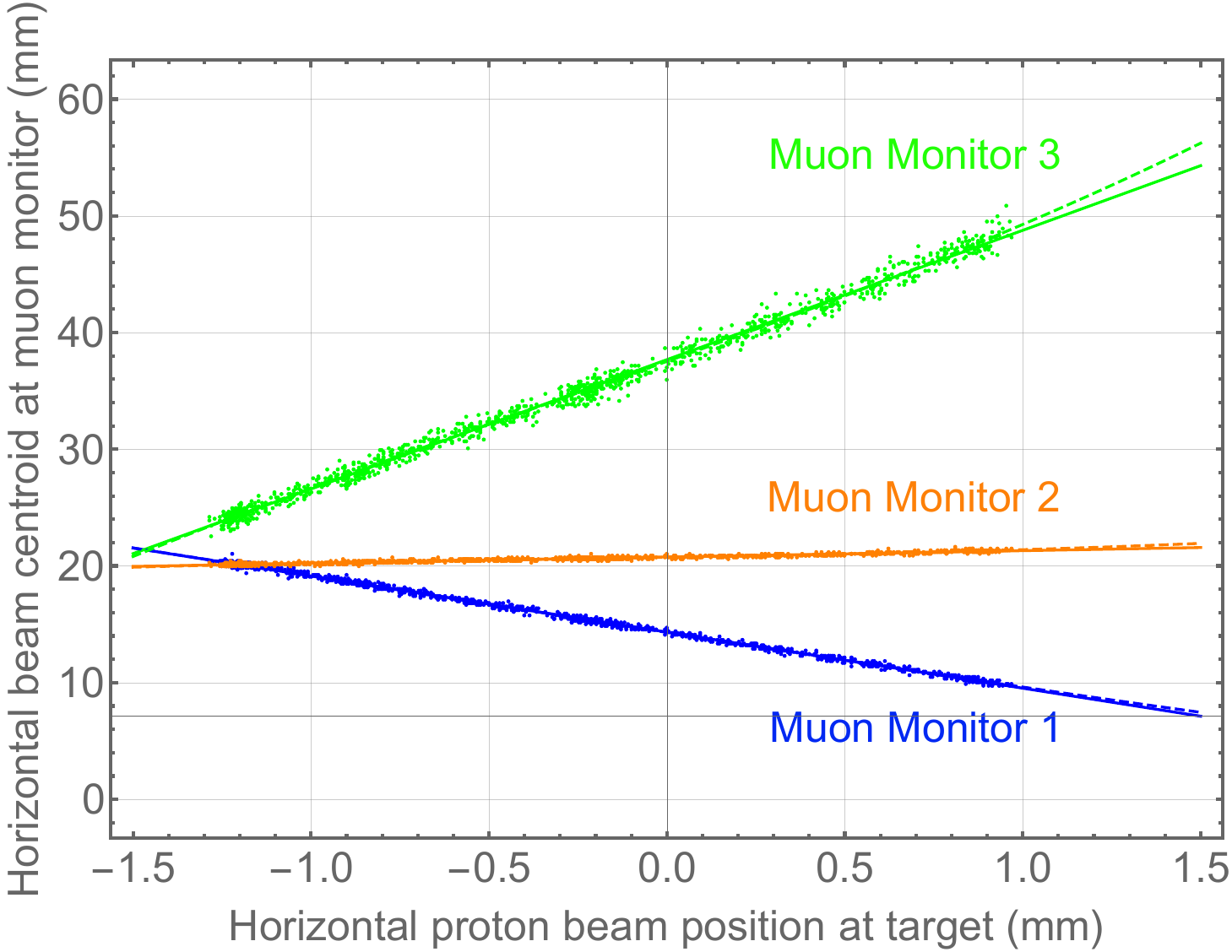}
\caption{
Horizontal muon beam centroids at the three muon monitors as a function of horizontal proton beam position on the target (horn current: $200$\,kA). 
Blue, orange, and green points represent Muon Monitor~1, 2, and 3, respectively. 
Solid and dashed lines show linear and third-order fits, respectively. 
}
\label{fig:200kahor}
\end{figure}

Figure~\ref{fig:200kahor} shows the horizontal muon beam centroid observed at each muon monitor as a function of the horizontal proton beam position on the target, with the horn current set to $200$\,kA.  
To evaluate the linearity of the response, two fitting functions are applied: a linear fit (solid line) and a third-order polynomial fit (dashed line).  
The two fits are nearly indistinguishable, confirming that the horizontal centroid variation is effectively linear.  
Variations in the intercepts among the monitors are attributed to imperfect alignment of the beam components and the monitors themselves relative to the beam axis. 
However, due to the wide muon profile, these offsets do not affect the measurement of relative centroid shifts. 
Alignment effects are not further analyzed in this study. 

The Muon Monitor~2 data are particularly noteworthy. 
The slope is nearly zero, indicating that the pions---and their resulting muons---are in focus at the monitor. 
This simple interpretation is consistent with the analysis presented in Sec.~\ref{sec:hornfocusingmechanism}. 
As shown in Fig.~\ref{fig:mm2pion}, the dominant contributing pions have transverse and longitudinal momenta  $p_{r,0} = 0.2$\,GeV/$c$ and $p_z = 18$\,GeV/$c$, respectively. 
Such pions yield $\lambda\sim 700$\,m, as shown in Fig.~\ref{fig:anavalid2}. 
Their corresponding incident radius $r$ at Horn~1 lies around $0.017$ and $0.02$\,m. 

The negative slope observed at Muon Monitor~1 indicates that the low-energy pions and muons are overfocused.
As a result, an upward displacement of the proton beam at the target causes the muon centroid at the monitor to shift downward, and vice versa, producing a negative correlation. 
This slope can also be interpreted as a magnification factor, analogous to image formation in light optics. 
The observed slope is approximately $-5$, implying the corresponding focusing parameter of $\lambda \sim 700$\,m $\times 1/6 \sim 120$\,m. 
According to Fig.~\ref{fig:anavalid1}, this corresponds to an initial radial position $r \sim 0.014$\,m, with an image size of approximately $70$\,mm radius on the muon monitor. 
In contrast, the positive slope observed at Muon Monitor~3 indicates that the pions and muons are underfocused or divergent at the monitor location. 

\begin{figure}
\includegraphics[width=8cm]{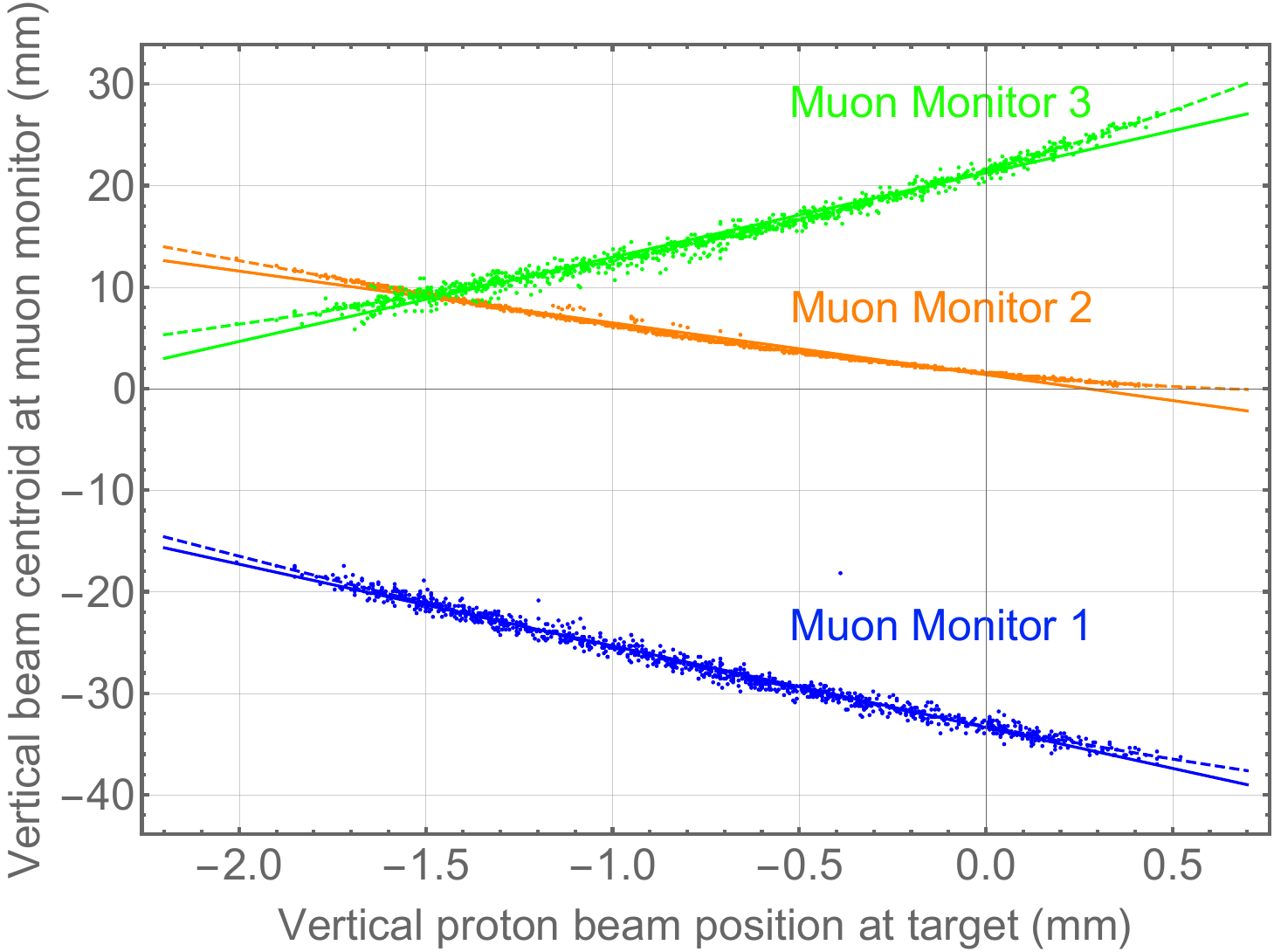}
\caption{
Vertical muon beam centroids at the three muon monitors as a function of the vertical proton beam position on the target (horn current: $200$\,kA). 
Blue, orange, and green points correspond to Muon Monitors~1, 2, and 3, respectively. 
Solid and dashed lines represent linear and third-order polynomial fits, respectively.}
\label{fig:200kaver}
\end{figure}

Figure~\ref{fig:200kaver} shows the vertical muon beam centroid at the three muon monitors as a function of the vertical proton beam position on the target, with the horn current set to 200\,kA.  
A third-order polynomial fit was applied to the data to capture any potential nonlinearities. 
The resulting best-fit curve indicates that the beam centroid variation remains predominantly linear, with only a small quadratic correction and negligible cubic contribution.
The slopes observed in the vertical scan are comparable to those in the horizontal scan, although the inferred radial position $r$ appears slightly larger in the vertical case. 
In particular, non-linear behavior becomes more pronounced in Muon Monitors~2 and 3. 
This is likely due to the vertical asymmetry in the target geometry, which causes the vertical pion phase space to differ from the horizontal case. 
Interestingly, a similar nonlinear response is observed in the full model simulation. 
However, the simulation does not perfectly reproduce the measured behavior, indicating potential areas for improvement---particularly in the modeling of hadronic interactions. 

\begin{figure}
\includegraphics[width=8cm]{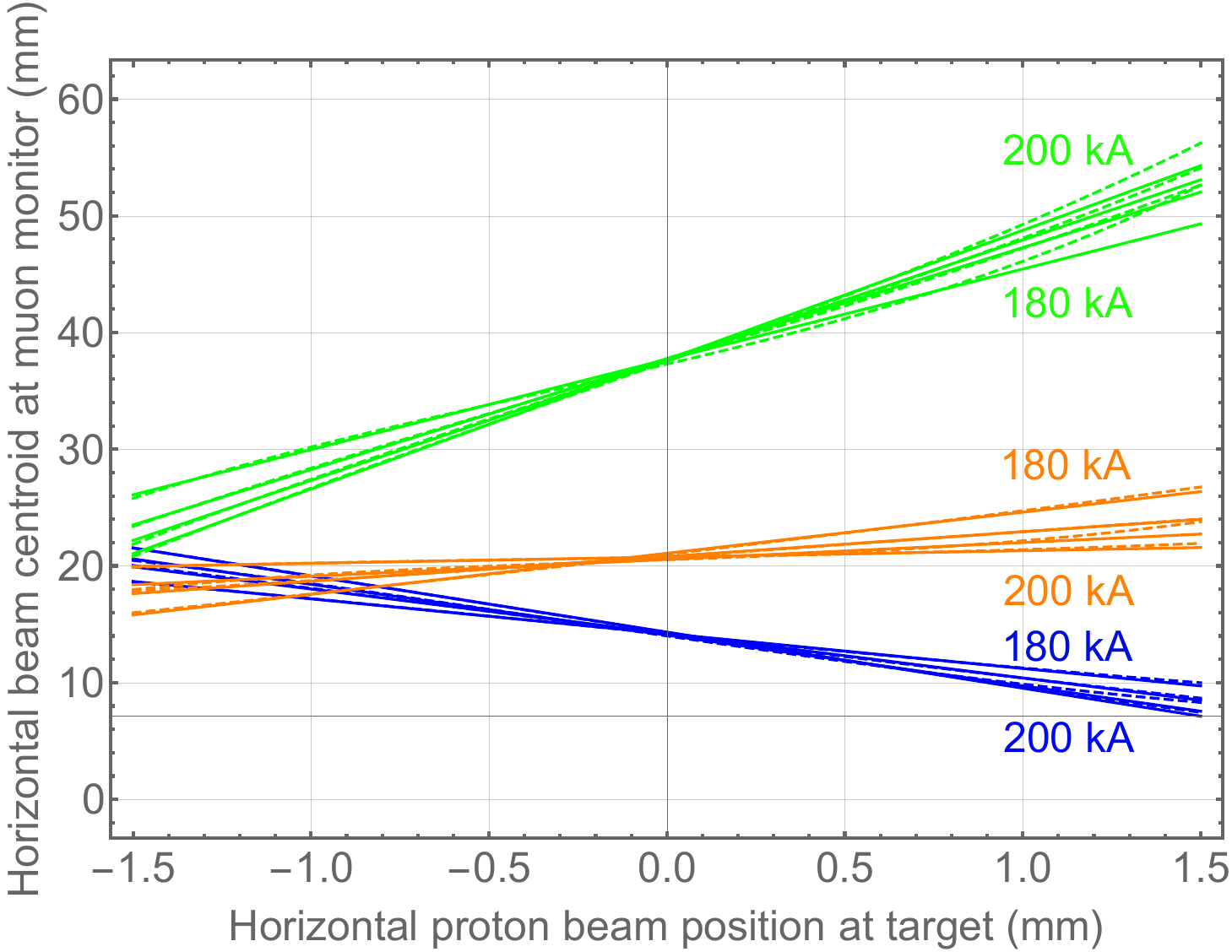}
\caption{
Fitted horizontal muon beam centroids at the three muon monitors as a function of  horizontal proton beam position on the target, for horn currents of $200$, $195$, $190$, and $180$\,kA. 
Blue, orange, and green curves correspond to Muon Monitors~1, 2, and 3, respectively. 
Solid lines are linear fits; dashed lines are third-order polynomial fits. 
}
\label{fig:xhorndep}
\end{figure}

\begin{figure}
\includegraphics[width=8cm]{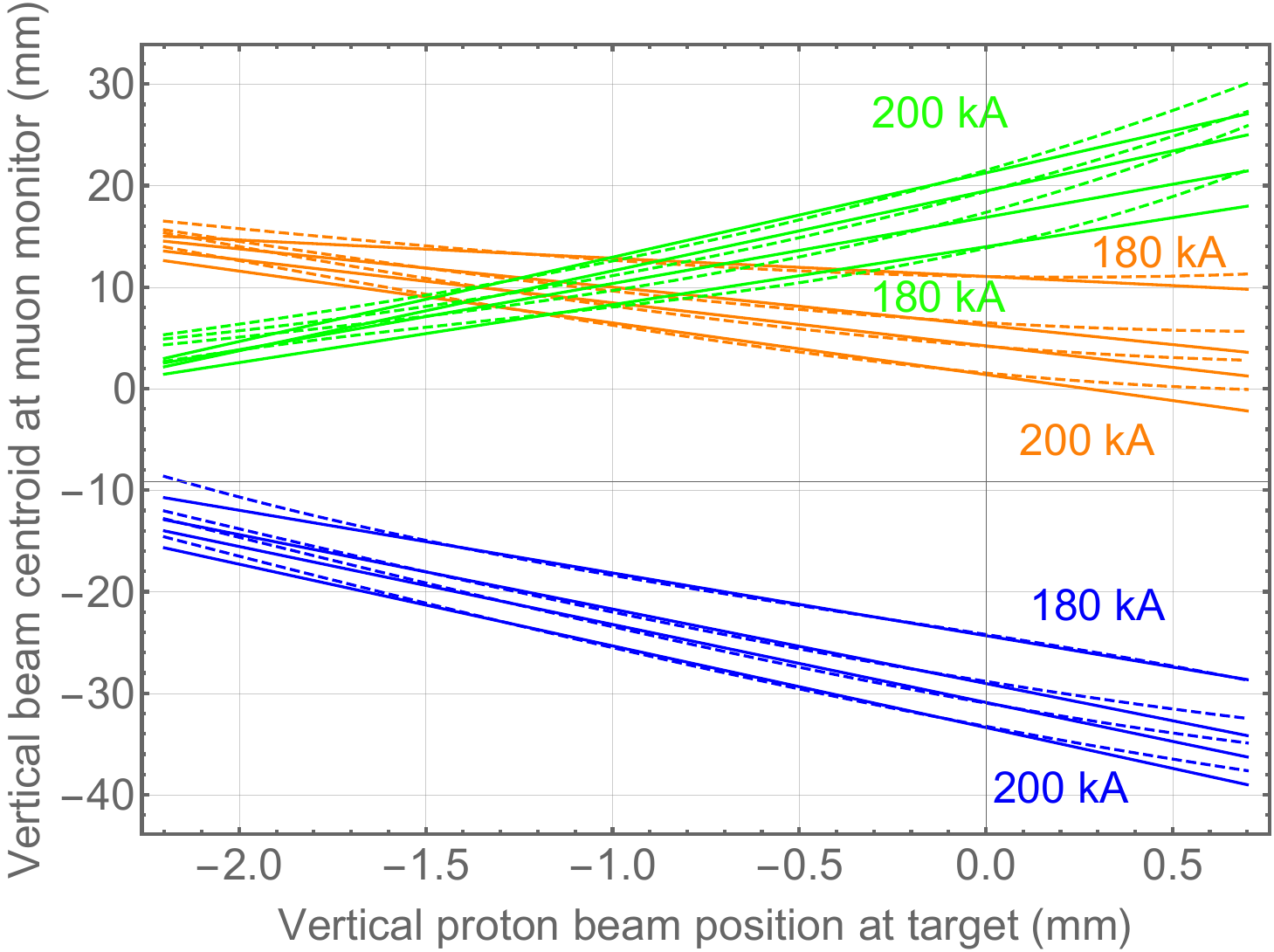}
\caption{
Fitted vertical muon beam centroids at the three muon monitors as a function of vertical proton beam position on the target, for horn currents of $200$, $195$, $190$, and $180$\,kA. 
Blue, orange, and green curves correspond to Muon Monitors~1, 2, and 3, respectively. 
Solid lines are linear fits;  dashed lines are third-order polynomial fits. 
}
\label{fig:yhorndep}
\end{figure}

\begin{figure}
\includegraphics[width=8cm]{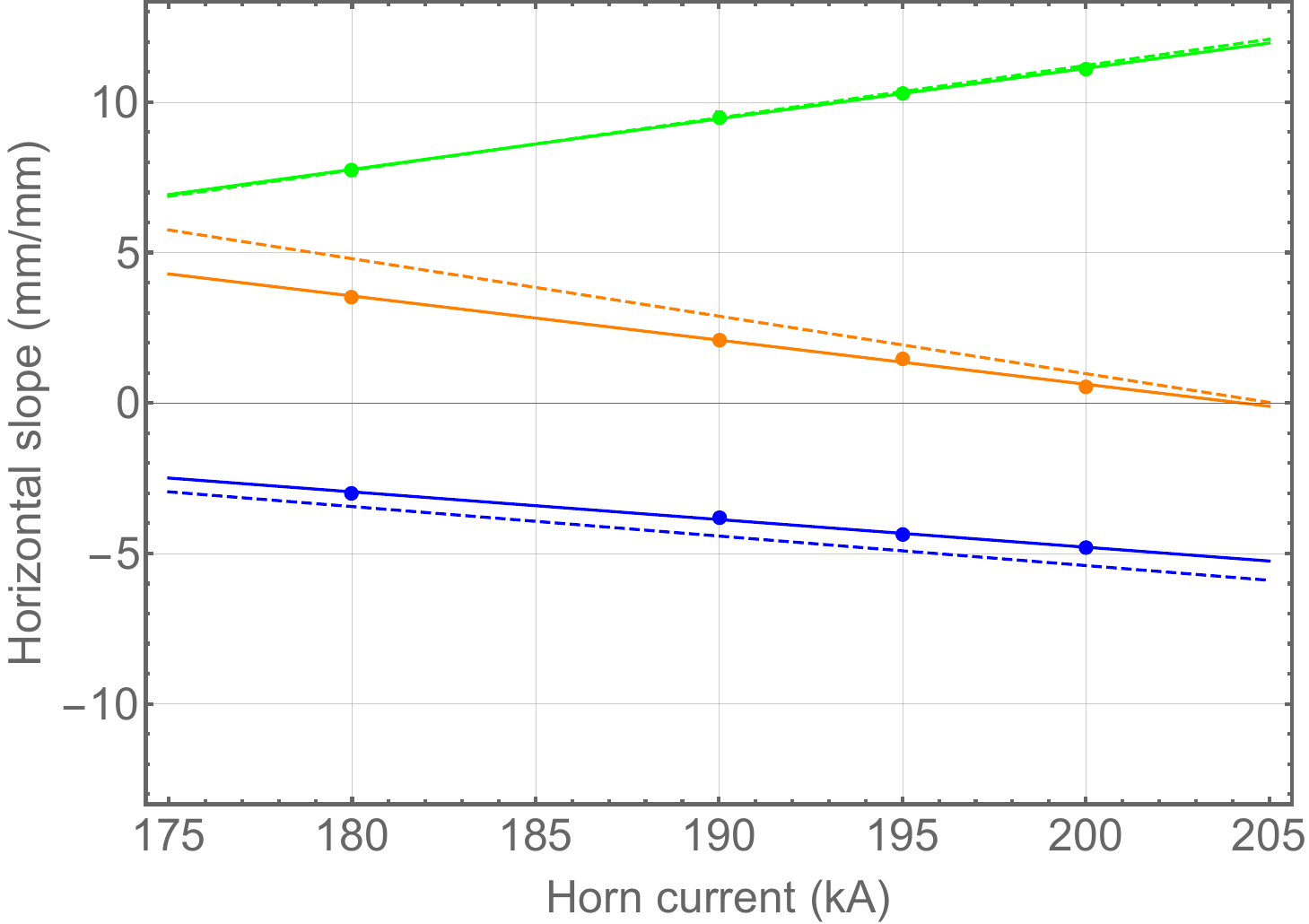}
\caption{
Measured slopes from the horizontal scan as a function of horn current. 
Blue, orange, and green points represent measured slopes at  Muon Monitors~1, 2, and 3, respectively. 
Dashed lines show corresponding predictions from the full model simulation.  
}
\label{fig:xslope}
\end{figure}

\begin{figure}
\includegraphics[width=8cm]{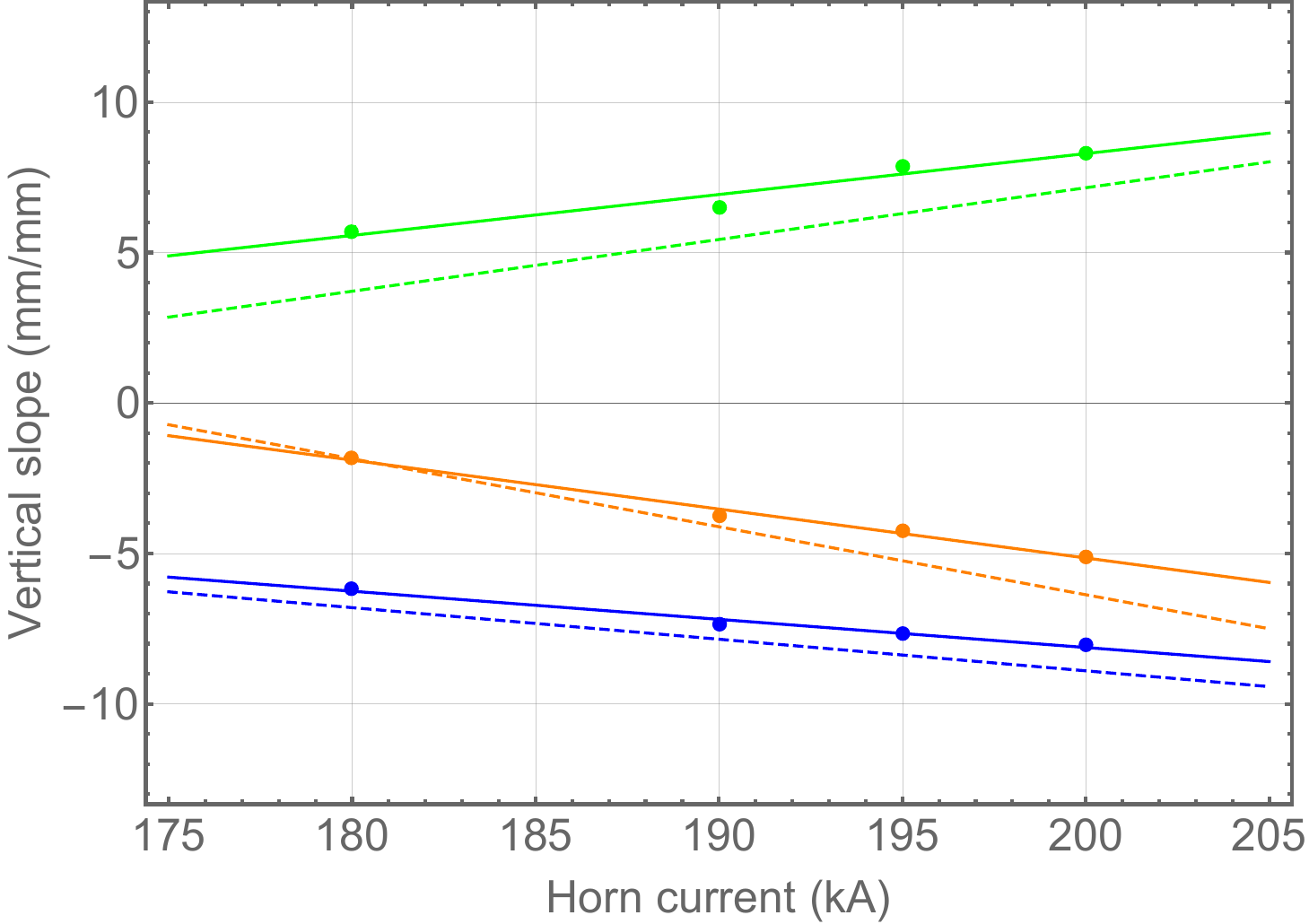}
\caption{
Measured slopes from the vertical scan as a function of horn current. 
Blue, orange, and green points represent measured slopes at Muon Monitor~1, 2, and 3. 
Dashed lines show predictions from the full model simulation. 
}
\label{fig:yslope}
\end{figure}

Figures~\ref{fig:xhorndep} and~\ref{fig:yhorndep} show the fitted horizontal and vertical muon beam centroids as a function of proton beam position for horn currents of $200$, $195$, $190$, and $180$\,kA, respectively.
Figures~\ref{fig:xslope} and~\ref{fig:yslope} show the measured slopes from horizontal and vertical beam scans as a function of horn current, respectively. 
Solid lines represent linear fits to the measured data, while dashed lines show predictions from the full model simulation. 
Both measurement and simulation indicate that the slope is proportional to the horn current, consistent with expectations from the analytical models, where the transverse kick is linear in current. 
Note that the slope variation reflects changes in the focusing strength and corresponding magnification factor. 

Overall, the full model simulation reproduces the measured trend well. 
However, it is important to note that the simulation does not include the delta-ray production~\cite{Loiacono2010}, which may influence the muon monitor signals. 
Although a dedicated delta-ray model is not currently enabled, the delta-ray yield is expected to be negligible for the present analysis. 
Additionally, potential effects from misalignment of beam components are not considered in this study and should be evaluated in future work. 
Another possible source of discrepancy is the imperfect modeling of the hadron interactions in the simulation. 
Extensive efforts by experiments such as NA49~\cite{NA49}, NA61~\cite{NA61}, HARP~\cite{HARP,HARP2}, MIPP~\cite{MIPP}, and EMPHATIC~\cite{EMPHATIC}, have significantly contributed to improving the accuracy of hadronic interaction models.

\section{Machine Learning Model for Detecting Beam Parameters using Muon Monitor Data}
\label{sec:machinelearning}
As shown in previous sections, the NuMI horn system exhibits predominantly linear optical behavior, implying a linear relationship between the muon monitor response and variations in beam parameters. 
To leverage this property, we apply a machine learning (ML) model to analyze the muon monitor signals. 
The advantage of using ML in this application lies in its ability to identify correlations among a large number of input variables and disentangle them to infer individual beam parameters. 
For example, it is known that the beam intensity can influence the horn current by heating the horn conductors, thereby altering their electrical conductivity. 
With ML analysis, the horn current and the beam intensity can be independently extracted from muon monitor data, despite their coupled effects on the observed profiles. 

An artificial neural network (ANN) with multiple hidden layers is used as the ML architecture. 
The network takes 241 input features, corresponding to spill-by-spill signal amplitudes from individual muon monitor channels.
A total of 13,563 data samples were prepared for this study. 
For model optimization, the dataset was randomly split into training ($70$\%) and validation ($30$\%) subsets.
Hyperparameter tuning was performed using Bayesian optimization to systematically search for the optimal model configuration. 
The key parameters---including learning rate, number of hidden layers, activation functions, and batch size---were tuned to minimize the standard error in model predictions. 
This approach significantly reduces the model's sensitivity to the specific ANN architecture, enabling accurate extraction of beam parameters. 
To mitigate the risk of biased predictions, data preprocessing included careful identification and removal of outliers. 
Outliers were flagged through correlation studies between muon monitor signals and known beam parameters, where deviations from expected trends indicated anomalies. 
Such anomalies, including faulty muon monitor channels, can introduce systematic uncertainties in the training data and impact model performance. 

\begin{figure}
\includegraphics[width=8cm]{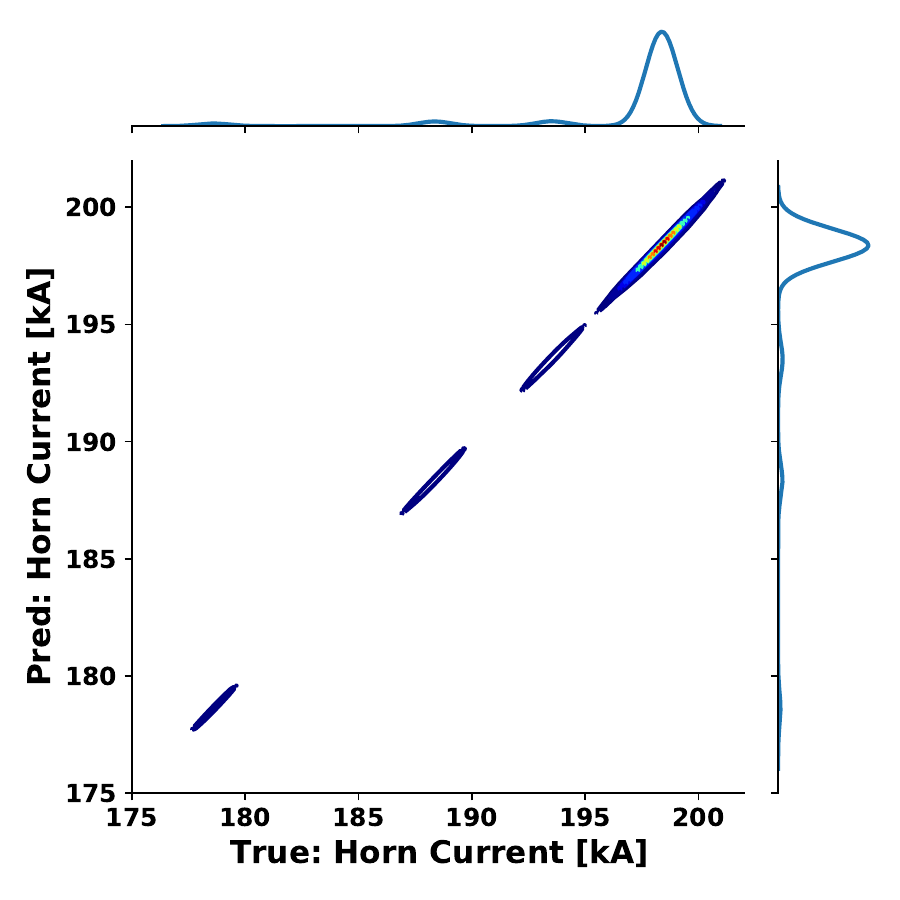}
\caption{Comparison between the horn current predicted from the muon monitor signals and the measured horn current by using the horn CT.}
\label{fig:mlhornrms}
\end{figure}

\begin{figure}
\includegraphics[width=8cm]{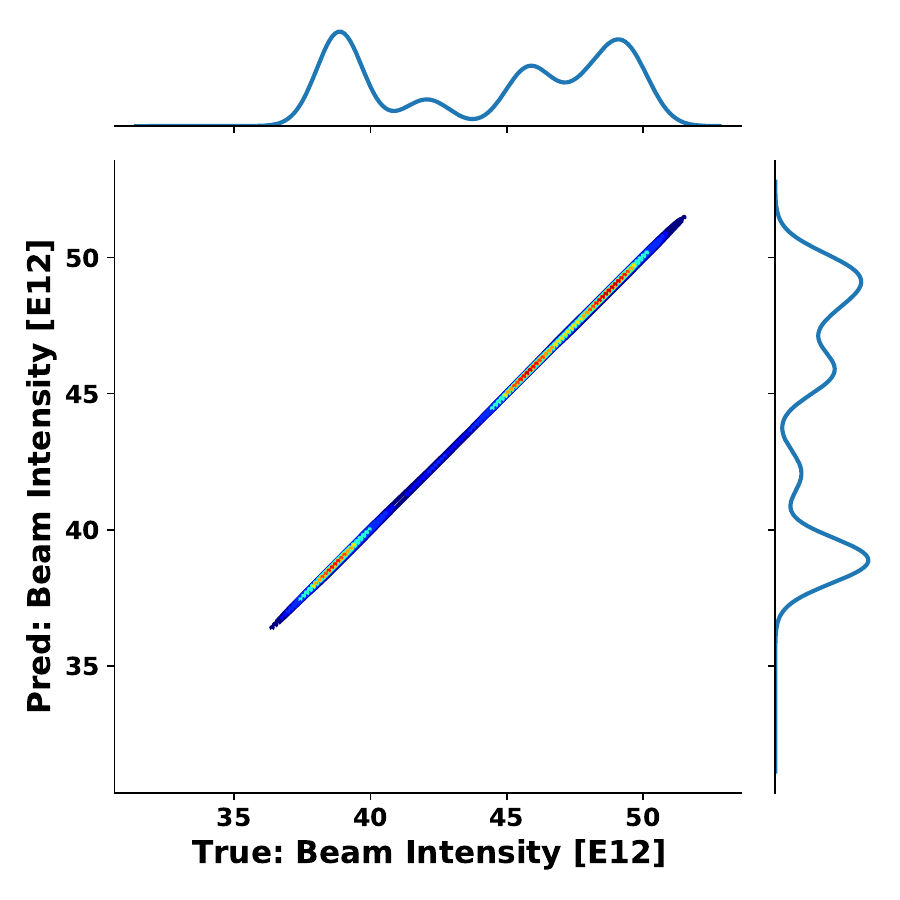}
\caption{
Comparison between the beam intensity predicted from the muon monitor signals and the measured intensity using the beam CT. 
The units are $10^{12}$ POT per spill.}
\label{fig:mlbeamint}
\end{figure}

Figure~\ref{fig:mlhornrms} compares the horn current measured by the horn current transformer (horn CT) with the value predicted by the trained ANN model using muon monitor signals as input. 
Similarly, 
Fig.~\ref{fig:mlbeamint} compares the proton beam intensity measured by the beam current transformer (beam CT) and the beam intensity predicted by the trained ANN model.
These results demonstrate that the trained ANN model can independently extract both the horn current and the beam intensity from the same validation data set. 
This confirms the ML model's ability to disentangle and accurately infer multiple beam parameters from the observed muon monitor profiles. 

\begin{figure}
\includegraphics[width=8cm]{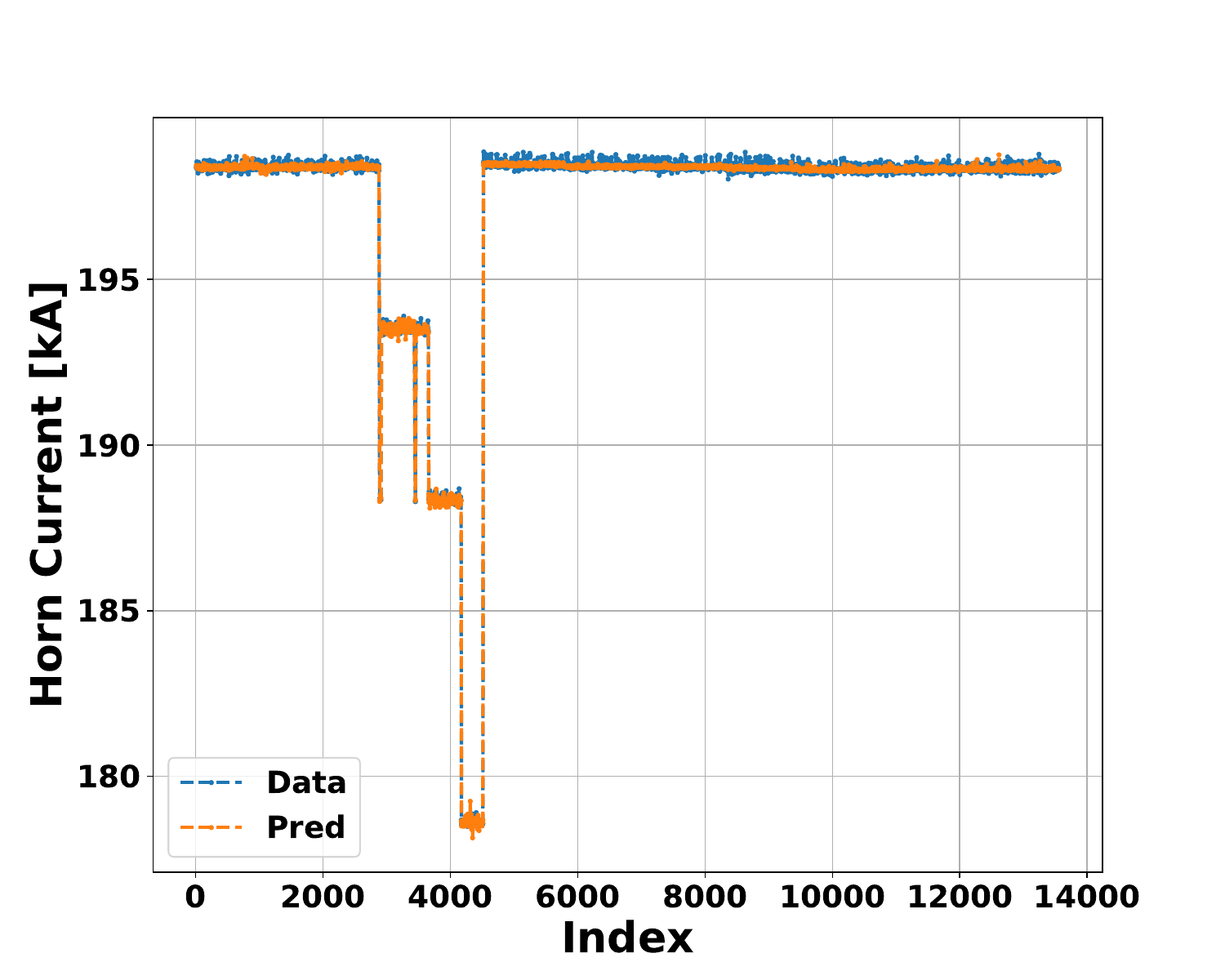}
\caption{
Comparison between the horn current measured by the horn CT and the predicted by the trained ANN using muon monitor signals.}
\label{fig:mlhornres}
\end{figure}
\begin{figure}
\includegraphics[width=8cm]{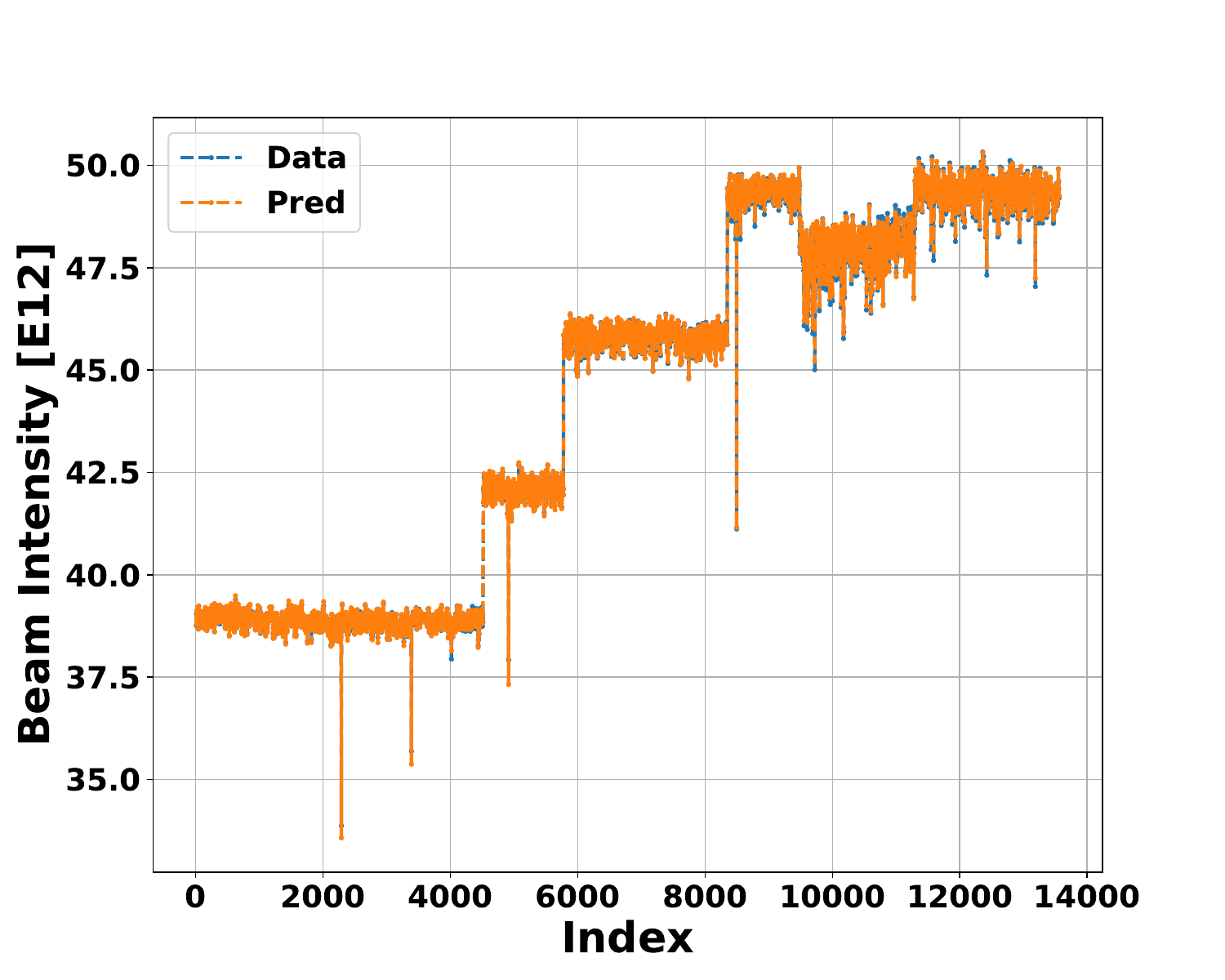}
\caption{
Comparison between the beam intensity measured by the beam CT and that predicted by the trained ANN model using muon monitor signals.}
\label{fig:mlbeamintres}
\end{figure}

Figures~\ref{fig:mlhornres} and~\ref{fig:mlbeamintres} present the measured and ANN-predicted values of horn current and beam intensity, respectively, as a function of the spill index. 
The model effectively captures the spill-by-spill fluctuations observed in the primary beam instrumentation signals, demonstrating its high temporal resolution and reliability. 
In this context, ``index" refers to the sequential spill number and serves as a proxy for time. 
These results suggest that the observed fluctuations in the primary beam instrumentation are due to real variations in beam conditions, rather than random electrical noise. 
The ANN model achieves high precision in estimating the beam parameters, with a precision of $\pm0.05$\% for the horn current and $\pm0.1$\% for the beam intensity. 

\begin{figure}
\includegraphics[width=8cm]{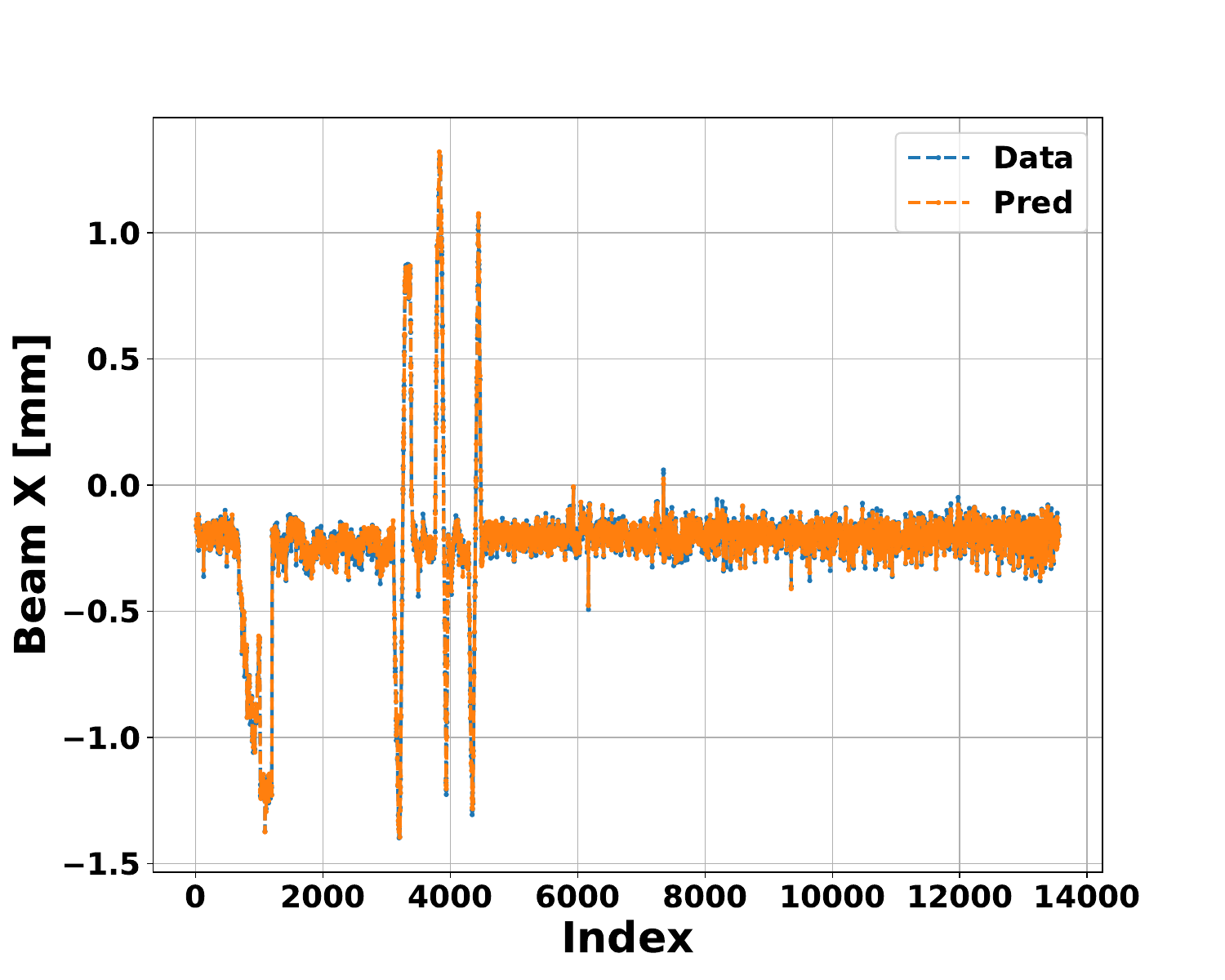}
\caption{Comparison between the horizontal beam position at the target, as measured by BPMs, and that predicted by the trained ANN model using muon monitor signals. 
}
\label{fig:mlbeamxres}
\end{figure}

\begin{figure}
\includegraphics[width=8cm]{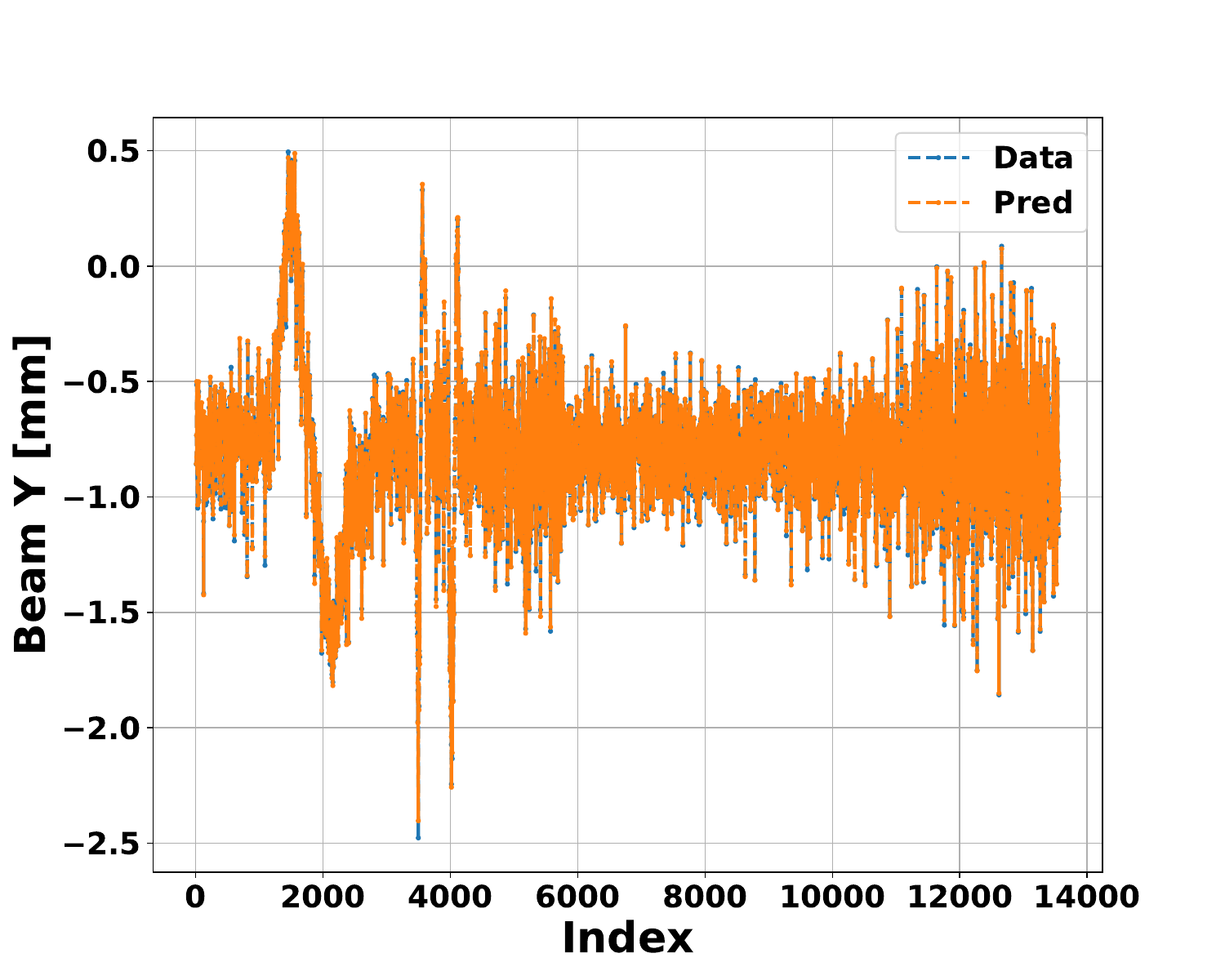}
\caption{Comparison between the vertical beam position at the target, as measured by BPMs, and that predicted by the trained ANN model using muon monitor signals. 
}
\label{fig:mlbeamyres}
\end{figure}

Figures~\ref{fig:mlbeamxres} and~\ref{fig:mlbeamyres} compare the horizontal and vertical beam positions at the target, measured by the beam position monitors (BPM), with the corresponding values predicted by the trained ANN model. 
The predicted beam positions closely match the measured data, indicating that the observed fluctuations are primarily due to actual variations in beam stability rather than random electrical noise.  
The precision of ML-based predictions is estimated to be $\pm0.018$\,mm in the horizontal direction and $\pm0.013$\,mm in the vertical direction. 

As summarized in Table~\ref{tab:beaminst}, the muon monitor, when combined with the trained ANN model, can effectively detect all critical beam parameters relevant to the NOvA experiment's physics tolerance.
The precision achieved is comparable to that of conventional primary beam instrumentation, as shown in Table~\ref{tab:phys}, demonstrating that the ML approach provides a reliable, noninvasive alternative for real-time beam monitoring.  

\begin{table}
\caption{\label{tab:beaminst}
Summary of beam parameters and the corresponding instrumentation used to monitor them. 
An open circle indicates parameters that can be reliably detected using the trained ML model; a triangle indicates parameters under evaluation for future facilities. 
}
\begin{ruledtabular}
\begin{tabular}{ lccccc } 
Beam parameter & BPM & PM & Beam CT & Horn CT & MM \\
 \hline
 Beam position & $\bigcirc$ & $\bigcirc$ & & & $\bigcirc$  \\ 
 Beam spot size & & $\bigcirc$ & & & $\triangle$ \\
 Beam intensity &
 & &  $\bigcirc$ & & $\bigcirc$ \\
 Horn current &
 & & & $\bigcirc$ & $\bigcirc$ \\
\end{tabular}
\end{ruledtabular}
\end{table}

These studies confirm that ML offers a strong potential to enhance diagnostic capabilities for the NuMI target and the horn system. 
One promising application is the use of ML as an anomaly detection tool. This tool can identify deviations such as reduced target density, damage to the target core, or degradation of the horn magnetic field quality.  
However, the accuracy of ML-based detection is strongly influenced by the quality of the training dataset. 
Optimizing the ML architecture and minimizing systematic uncertainties in the training data are essential to improving model performance. 
Ongoing efforts continue to address these areas of development. 

\section{Conclusion and Future Prospects}
\label{sec:summary}
This study investigated the focusing behavior of the NuMI magnetic horn system and demonstrated the feasibility of extracting key beam parameters using the downstream muon monitor data. 
By modeling the horn optics with an analytically derived single-horn framework and validating it through semi-analytical and numerical simulations, we confirmed that the NuMI horn system exhibits predominantly linear focusing characteristics. 
This linearity establishes a direct and predictable relationship between the incoming proton beam and the resulting pion and muon phase space distributions.

Building on this framework, we showed that essential beam parameters---including proton beam intensity, horn current, and beam position on target---can be accurately reconstructed using machine learning techniques applied to muon monitor observations. 
The trained artificial neural network (ANN) model achieved high precision, detecting horn current variations to within $\pm0.05$\% and beam position determining the beam position at the target with precision of $\pm0.018$\,mm (horizontal) and $\pm0.013$\,mm (vertical)---comparable to conventional primary beamline instrumentation.



These results demonstrate that muon monitors can serve as effective secondary diagnostics, providing a robust cross-check for beam conditions and helping to reduce systematic uncertainties. 
This capability is particularly relevant for future long-baseline experiments such as the Deep Underground Neutrino Experiment (DUNE), which requires high-precision control and monitoring of the neutrino beam delivered by the Long-Baseline Neutrino Facility (LBNF).

Further refinement of full-model simulations is needed to capture residual discrepancies in muon profiles, particularly those arising from delta-ray backgrounds and beamline component misalignment. 
Improved modeling of hadronic interactions will also be critical for achieving the accuracy required in next-generation neutrino experiments.
This work underscores the value of combining analytical modeling, simulation, and machine learning to improve beam diagnostics. 
As precision demands continue to grow in neutrino physics, these methods will play a central role in enabling accurate and reliable measurements for the next generation of high-intensity accelerator-based experiments.


\appendix
\section{Particle Motion in Horn Magnets}
\label{sec:appendix_analytical}
We first develop an analytical formulation to describe the motion of charged particles in the magnetic field generated by an infinitely long, thin, straight wire. 
We then derive the phase space evolution and perform numerical simulations to validate the model. 
Finally, we introduce an approximation to analytically describe a single-horn model used to analyze the NuMI horn focusing mechanism.

\subsection{Analytical Formulation of Particle Motion in Magnetic Field Generated by an Infinitely Long Thin Straight Wire}
The analytical formulation is originally described in Ref.~\cite{Regenstreif}. 
Particle motion is represented in a cylindrical coordinate system $(r,\phi,z)$. 
The general equations of motion are, 
\begin{eqnarray}
m \left(\ddot{r}-r \dot{\phi}^2 \right)=&q \left( r\dot{\phi} b_z -\dot{z} b_{\phi} \right),
\label{eq:radial}\\
\frac{m}{r}\frac{d}{dt}\left( r^2 \dot{\phi}\right) =&q \left( \dot{z} b_r - \dot{r} b_z \right),
\label{eq:azimuthal}\\
m \ddot{z} =& q \left(\dot{r} b_{\phi}-r \dot{\phi} b_r \right),
\label{eq:longitudinal}
\end{eqnarray}
where $m=\gamma m_0$ is the relativistic mass, with $\gamma$ being the Lorentz factor and $m_0$ the rest mass. 
$q$ is the particle charge. 
The magnetic field components in cylindrical coordinates are denoted as
$b_r$, $b_\phi$ and $b_z$. 
The total particle velocity is given by
\begin{equation}
v^2 = \dot{r}^2+r^2 \dot{\phi}^2 + \dot{z}^2. 
\label{eq:velocity}
\end{equation}
For a magnetic field generated by an infinitely long, thin, straight wire carrying current $I$, the field is purely azimuthal, 
\begin{equation}
b_z=b_r=0, b_\phi=\frac{\mu_0 I}{2 \pi r}, 
\label{eq:approximation1}
\end{equation}
where $\mu_0$ is the permeability of free space. 
Substituting this into Eq.~(\ref{eq:azimuthal}) yields the conservation of angular momentum,  
\begin{equation}
r^2 \dot{\phi}=r_0^2 \dot{\phi_0}=C=\textrm{constant}.
\label{eq:angularmomentum}
\end{equation}
Using this result, the radial and longitudinal equations of motion become,
\begin{eqnarray}
\ddot{r}-\frac{C^2}{r^3}=&-\frac{\mu_0 I q}{2 \pi m}\frac{\dot{z}}{r},
\label{eq:radial2}\\
\ddot{z}=&\frac{\mu_0 I q}{2 \pi m}\frac{\dot{r}}{r}. 
\label{eq:azimuthal2}
\end{eqnarray}
Defining a constant $V=\frac{\mu_0 Iq}{2 \pi m}$, Eqs.~(\ref{eq:radial2}) and (\ref{eq:azimuthal2})  simplify to, 
\begin{eqnarray}
\ddot{r}-\frac{C^2}{r^3}=&-V\frac{\dot{z}}{r},\\
\label{eq:radial3}
\ddot{z}=&V\frac{\dot{r}}{r}. 
\label{eq:azimuthal3}
\end{eqnarray}
To express these equations in terms of the longitudinal coordinate $z$, we use, 
\begin{eqnarray}
\dot{r}=\frac{dr}{dt}=\frac{dz}{dt}\frac{dr}{dz}=r'\dot{z}, 
\label{eq:radsimple1}\\
\ddot{r}=\frac{d^2r}{dt^2}=r'\ddot{z}+r''\dot{z}^2. 
\label{eq:radsimple2}
\end{eqnarray}
Substituting Eqs.~(\ref{eq:angularmomentum}), (\ref{eq:radsimple1}) and (\ref{eq:radsimple2}) into Eq.(\ref{eq:velocity}) and simplifying the longitudinal velocity yields
\begin{eqnarray}
\dot{z}=\pm \sqrt{\frac{v^2-\frac{C^2}{r^2}}{1+r'^2}}. 
\label{eq:zdot}
\end{eqnarray}
Combining the expressions for $\dot{r}$, $\ddot{r}$, and $\dot{z}$, we derive a second-order radial differential equation, 
\begin{equation}
\begin{split}
&r'' \left(v^2 - \frac{C^2}{r^2} \right) \\
&- \frac{C^2}{r^3} \left(1 + r'^2 \right)^{3/2}
\sqrt{v^2 - \frac{C^2}{r^2}} = 0
\label{eq:radial4}
\end{split}
\end{equation}
This can be reduced to a first-order differential equation using the relation, 
\begin{eqnarray}
d \left( \frac{\sqrt{v^2-\frac{C^2}{r^2}}}{1+r'^2} \right) = \pm V\frac{dr}{r}.
\label{eq:radial5}
\end{eqnarray}
Eq.~(\ref{eq:radial4}) is transformed into
\begin{eqnarray}
\frac{\sqrt{v^2-\frac{C^2}{r^2}}}{1+r'^2}=\pm V \ln (r) + \textrm{constant}.
\label{eq:radial6}
\end{eqnarray}
Applying an initial condition $r_0'$ at $r_0$, Eq.(\ref{eq:radial6}) becomes,
\begin{eqnarray}
\frac{\sqrt{v^2-\frac{C^2}{r^2}}}{1+r'^2}=\pm V \ln \left(\frac{r}{r_0} \right)+
\frac{\sqrt{v^2-\frac{C^2}{r_0^2}}}{1+r_0'^2}.
\label{eq:radial7}
\end{eqnarray}

Following a similar procedure for azimuthal motion, we obtain, 
\begin{align}
    &\dot{\phi} =\dot{z} \frac{d\phi}{dz} = \dot{z} \frac{d\phi}{dr} r', \\
    \begin{split}
    &\sqrt{v^2 - \frac{C^2}{r^2} \left(1 + \frac{1}{r^2} \left(\frac{dr}{d\phi}\right)^2 \right)}\\ &= \pm V \ln \left( \frac{r}{r_0} \right) 
    + \frac{\sqrt{v^2 - \frac{C^2}{r_0^2}}}{1 + r_0'^2}. 
    \end{split}
    \label{eq:azimuthal7}
\end{align}

To simplify the description of the trajectory geometry, we define geometric transformation parameters, 
\begin{eqnarray}
\tan \beta = r', \tan \alpha=\phi', \tan \gamma=\cos \beta \cdot \tan \alpha. 
\label{eq:conversion}
\end{eqnarray}
\begin{figure}
\includegraphics[width=8cm]{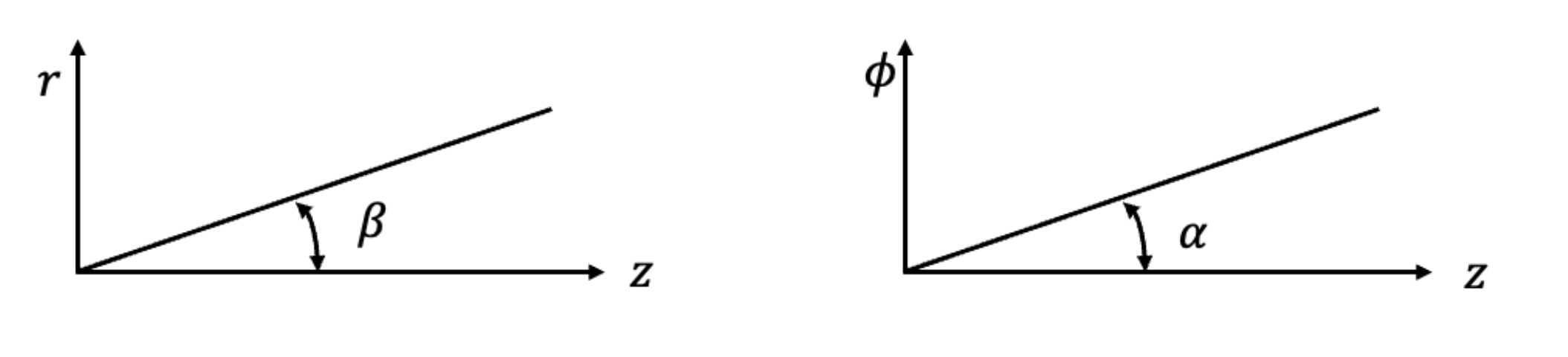}
\caption{Translate $r'$ and $\phi'$ into $\beta$ and $\alpha$.
}
\label{fig:geometry_constraint}
\end{figure}
Using $\beta$ and $\alpha$, 
\begin{align}
    v^2 &= \left( 1+ \tan^2 \beta + \tan^2 \alpha \right) \dot{z}^2, \label{eq:velocity2} \\
    \dot{z}^2 &= \left( v^2 - \frac{C^2}{r^2} \right) \cos^2 \beta, \label{eq:velocity3} \\
    \begin{split}
    \frac{C^2}{r^2} &= v^2 \frac{\cos^2\beta \cdot \tan^2\alpha}{1 + \cos^2\beta \cdot \tan^2\alpha} \\
    &= v^2 \frac{\tan^2\gamma}{1 + \tan^2\gamma} = v^2 \sin^2\gamma. 
    \end{split}
    \label{eq:velocity4}
\end{align}

\begin{figure}
\includegraphics[width=8cm]{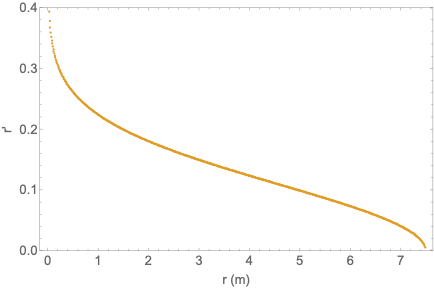}
\caption{Simulated phase space evolution in an infinitely long thin straight wire. 
}
\label{fig:phasespace1}
\end{figure}

Using the geometric transformations defined above, we additionally introduce the dimensionless parameters $x = r/r_0$ and $\hat\lambda = V/v$ to simplify the final equations, 
\begin{align}
    x &= \frac{\sin \gamma_0}{\sin \gamma}, \label{eq:finaleq1} \\
    \begin{split}
    &\left( 1 - \frac{\sin^2\gamma_0}{x^2} \right) \left[ 1 + r_0^2 \left( \frac{dx}{dz} \right)^2 \right]^{-1} \\
    &= \left( \cos\beta_0 \cos\gamma_0 \pm \hat\lambda \ln x \right)^2, 
    \end{split}
    \label{eq:finaleq2} \\
    \begin{split}
    &1 - \frac{\sin^2\gamma_0}{x^2} \left[ 1 + \frac{1}{x^2} \left( \frac{dx}{d\phi} \right)^2 \right] \\
    &= \left( \cos\beta_0 \cos\gamma_0 \pm \hat\lambda \ln x \right)^2.
    \end{split}
    \label{eq:finaleq3}
\end{align}
In contrast to the full treatment in Ref.~\cite{Regenstreif}, we focus on the radial dynamics described by Eq.~(\ref{eq:finaleq2}), which provides the phase space $(r, r')$. 
Fig.~\ref{fig:phasespace1} shows the evolution of the simulated phase space evolution obtained by numerically solving Eq.~(\ref{eq:finaleq2}). 
In the simulation, the wire current is set to $200$\,kA. 
The test particle is a pion with a momentum of $1$\,GeV/$c$, initially located at $r_0 = 10$\,mm, with an initial injection angle $r'_0 =\phi'_0 = 0.3$\,rad. 
Initially, the particle moves outward from the wire, and as it reaches its maximum radial position, its trajectory becomes nearly parallel to the axis. 
It then curves inward again under the influence of the azimuthal magnetic field, which increases as the particle approaches the wire. 
This process repeats, resulting in periodic behavior in the phase space.

\begin{figure}
\includegraphics[width=8cm]{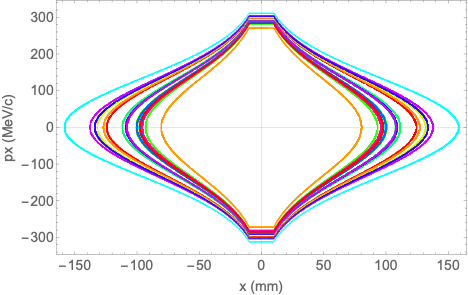}
\caption{
Simulated phase space evolution in an infinitely long coaxial inner conductor magnet. 
The radius of the conductor is $10$\,mm and a current is set to $300$\,kA. 
There are 11 particles with their average initial angle $0.15$\,rad with the distributed RMS angle $0.01$\,rad. 
They follows the same trajectory for each revolution.
}
\label{fig:phasespace3}
\end{figure}

Since Eq.~(\ref{eq:finaleq2}) cannot be solved analytically, we rely on a numerical tracking to analyze the phase space behavior, rather than extracting a transfer matrix. 
In the analytical expression, $b_\phi$ is infinite at $r = 0$. 
To avoid such a singularity in the numerical simulation, a coaxial inner conductor is modeled as a field generator.  
Figure~\ref{fig:phasespace3} shows the resulting phase space evolution from the particle tracking simulation, with the coaxial inner conductor in place. 
All particles closely follow the same phase space trajectory with each oscillation, confirming the stability and periodicity of the motion predicted by Eq.~(\ref{eq:finaleq2}). 

\subsection{Simplified Equation for Radial Motion}
Because the neutrino horn has a finite length and a parabolic inner conductor profile, we introduce an approximation to simplify the equations of motion under this geometric constraint. 
We begin by neglecting the angular momentum term, effectively setting $C\rightarrow0$. 
The equations then reduce to, 
\begin{eqnarray}
\ddot{r}=&-V\frac{\dot{z}}{r}, 
\label{eq:simple0}\\
\ddot{z}=&V\frac{\dot{r}}{r}. 
\label{eq:simple1}
\end{eqnarray}
Substituting $\dot{r}=\dot{z}r'$ into Eq.(\ref{eq:simple1}) gives, 
\begin{eqnarray}
\ddot{z}=V\frac{\dot{z}}{r}r'=-r'\ddot{r}. 
\end{eqnarray}
This result shows that $\ddot{z}$ is proportional to $\ddot{r}$ through the radial slope $r'$. 
Since $r' \sim p_r/p_z = 0.3/p_z$ is a good approximation for neutrino target applications and $p_z$ for most pions---resulting muons reach to the muon monitors---is above $5$\,GeV/$c$ (shown in Fig.~$\ref{fig:mm1pion}$),
this implies that $r' < 0.3/5 = 0.06$\,rad. 
Therefore, $\ddot{z}\sim 0$ is a good approximation (paraxial approximation), 
\begin{eqnarray}
\ddot{r}=\frac{1}{m}\frac{dp_r}{dt}=\frac{v_z}{m}p_r',\\
\dot{z}=\frac{1}{m}p_{z,0},
\end{eqnarray}
where $p_{z,0}$ is constant. 
Substituting these into the radial equation, we find, 
\begin{eqnarray}
p_r'=&-\frac{\mu_0 q I}{2 \pi mv_z}\frac{p_{z,0}}{r}, \\
dp_r=&-\frac{\mu_0 q I}{2 \pi r}dz, \\
p_r=&p_{r,0}-\int \frac{\mu_0 q I}{2 \pi r}dz. 
\label{eq:simple2}
\end{eqnarray}
A key principle of horn focusing is that the particle's path length within the magnetic field region follows a parabolic profile, reflecting the shape of the inner conductor. 
This is approximated by substituting $\int dz  \rightarrow a\cdot r^2$, where $a$ is a parabolic profile coefficient. 
Dividing both sides of the equation by $p_{z,0}$ and 
using $p_r/p_{z,0} \sim \theta$ and $p_{r,0}/p_{z,0} \sim \theta_0$, 
we obtain a simplified expression, 
\begin{eqnarray}\
\frac{p_r}{p_{z,0}}\sim \theta = \theta_0-\frac{\mu_0 q I}{2 \pi p_{z,0}}ar. 
\label{eq:simple3}
\end{eqnarray}
This expression represents the fundamental relationship between the initial and final transverse angles of a particle as it traverse the horn field. 
The condition for perfect focusing, $\theta=0$, corresponds to, 
\begin{eqnarray}
\theta_0=\frac{\mu_0 q I}{2 \pi p_{z,0}}ar_0, \\
f=\frac{r_0}{\tan\theta_0}\sim \frac{2 \pi p_{z,0}}{\mu_0 q I a}.
\label{eq:focuscondition}
\end{eqnarray}

\begin{acknowledgments}
We thank Fermilab Accelerator Operations, Target Systems, External Beam Delivery, and Main Injector Departments for their contributions to this study. 
We also thank the NOvA Collaboration for giving us the opportunity to run the special beam tests. 

This work was produced by Fermi Forward Discovery Group, LLC under Contract No. 89243024CSC000002 with the U.S. Department of Energy, Office of Science, Office of High Energy Physics. Publisher acknowledges the U.S. Government license to provide public access under the DOE Public Access Plan \url{https://www.energy.gov/doe-public-access-plan}.


\end{acknowledgments}

\bibliography{horn}

\end{document}